\documentclass[twocolumn,preprintnumbers,amsmath,amssymb,aps,nofootinbib]{revtex4}

\usepackage{graphicx}
\usepackage{dcolumn}
\usepackage{bm}
\usepackage{amsfonts,fancybox,ascmac,amsmath}
\usepackage{amsthm}
\usepackage{amssymb}
\usepackage{mathrsfs}

\newcommand{\Image}{{\rm Im}\hspace{0.07cm}}
\newcommand{\Kernel}{{\rm Ker}\hspace{0.07cm}}

\begin{document}

\title{Geometric control theory for quantum back-action evasion}
\author{Yu Yokotera}%
\email[E-mail address: ]{y-yokotera@z6.keio.jp}
\author{Naoki Yamamoto}
\email[E-mail address: ]{yamamoto@appi.keio.ac.jp}
\affiliation{Department of Applied Physics and Physico-Informatics, Keio University, Yokohama 223-8522, Japan}
\date{\today}%


\begin{abstract}
Engineering a sensor system for detecting an extremely tiny signal such as the 
gravitational-wave force is a very important subject in quantum physics. 
A major obstacle to this goal is that, in a simple detection setup, the measurement 
noise is lower bounded by the so-called standard quantum limit (SQL), which is 
originated from the intrinsic mechanical back-action noise. 
Hence, the sensor system has to be carefully engineered so that it evades 
the back-action noise and eventually beats the SQL. 
In this paper, based on the well-developed geometric control theory for classical 
disturbance decoupling problem, we provide a general method for designing 
an auxiliary (coherent feedback or direct interaction) controller for the sensor 
system to achieve the above-mentioned goal. 
This general theory is applied to a typical opto-mechanical sensor system. 
Also, we demonstrate a controller design for a practical situation where 
several experimental imperfections are present. 
\end{abstract}


\maketitle

\section{Introduction}
\label{sec:1}

Detecting a very weak signal which is almost inaccessible within the 
classical (i.e., non-quantum) regime is one of the most important subjects 
in quantum information science. 
A strong motivation to devise such an ultra-precise sensor stems from 
the field of gravitational wave detection 
\cite{BraginskyBook, Caves1980, Braginsky2008, Miao2012, Abbott2016}. 
In fact, a variety of linear sensors composed of opto-mechanical oscillators 
have been proposed \cite{Milburn,Chen2013, Latune2013, Aspelmeyer2014}, 
and several experimental implementations of those systems in various scales 
have been reported 
\cite{Corbitt2007, Matsumoto2015, Underwood2015, Thompson2008, Verhagen2012}.

It is well known that in general a linear sensor is subjected to two types of 
fundamental noises, i.e., the {\it back-action noise} and the {\it shot noise}. 
As a consequence, the measurement noise is lower bounded by the {\it standard 
quantum limit (SQL)} \cite{BraginskyBook, Caves1980}, which is mainly due 
to the presence of back-action noise. 
Hence, high-precision detection of a weak signal requires us to devise a sensor 
that evades the back-action noise and eventually beats the SQL; 
i.e., we need to have a sensor achieving {\it back-action evasion (BAE)}. 
In fact, many BAE methods have been developed especially in the field of 
gravitational wave detection, e.g., the variational measurement technique 
\cite{Vyatchanin1995, Kimble2001, Khalili2007} or the quantum locking scheme 
\cite{CourtyEurophys2003, CourtyPhysRev2003, Vitali2004}.  
Moreover, towards more accurate detection, recently we find some high-level 
approaches to design a BAE sensor, based on those specific BAE methods. 
For instance, Ref.~\cite{Miao2014} provides a systematic comparison of 
several BAE methods and gives an optimal solution. 
Also systems and control theoretical methods have been developed to synthesize 
a BAE sensor for a specific opto-mechanical system 
\cite{Tsang2010, YamamotoCFvsMF2014}; 
in particular, the synthesis is conducted by connecting an auxiliary system 
to a given plant system by {\it direct-interaction} \cite{Tsang2010} or 
{\it coherent feedback} \cite{YamamotoCFvsMF2014}.

Along this research direction, therefore, in this paper we set the goal to develop 
a general systems and control theory for engineering a sensor achieving BAE, 
for both the coherent feedback and the direct-interaction configurations. 
The key tool used here is the {\it geometric control theory} 
\cite{Schumacher1980, Schumacher1982, Wonham1985, Marro2010, Otsuka2015}, 
which had been developed a long time ago. 
This is indeed a beautiful theory providing a variety of controller design 
methods for various purposes such as the non-interacting control and 
the disturbance decoupling problem, but, to our best knowledge, it has not been 
applied to problems in quantum physics. 
Actually in this paper we first demonstrate that the general synthesis problem of 
a BAE sensor can be formulated and solved within the framework of geometric control theory, 
particularly the above-mentioned disturbance decoupling problem. \vspace{0.7mm}\\
\hspace{0.2cm}
This paper is organized as follows. 
Section~\ref{sec:2} is devoted to some preliminaries including a review of the 
geometric control theory, the general model of linear quantum systems, and 
the idea of BAE. 
Then, in Section~\ref{sec:3}, we provide the general theory for designing a 
coherent feedback controller achieving BAE, and demonstrate an example for 
an opto-mechanical system. 
In Section~\ref{sec:4}, we discuss the case of direct interaction scheme, also 
based on the geometric control theory. 
Finally, in Section~\ref{sec:5}, for a realistic opto-mechanical system subjected 
to a thermal environment (the perfect BAE is impossible in this case), we provide 
a convenient method to find an approximated BAE controller and show how 
much the designed controller can suppress the noise. \vspace{0.7mm}\\
\hspace{0.2cm}
{\bf Notation:} 
For a matrix $A=(a_{ij})$, $A^{\top}$, $A^{\dag}$, 
and $A^{\sharp}$ represent the transpose, Hermitian conjugate, 
and element-wise complex conjugate of $A$, respectively; i.e., 
$A^{\top}=(a_{ji})$,  $A^{\dag}=(a_{ji}^*)$, and 
$A^{\sharp}=(a_{ij}^*)=(A^{\dag})^{\top}$. 
$\Re(a)$ and $\Im(a)$ denote the real and imaginary parts of a complex 
number $a$. 
$O$ and $I_{n}$ denote the zero matrix and the $n \times n$ identity matrix. 
$\Kernel A$ and $\Image A$ denote the kernel and the image of 
a matrix $A$, i.e., $\Kernel A=\{x\,|\, Ax=0\}$ and 
$\Image A=\{y\,|\, y=Ax, ~\forall x\}$.


\section{Preliminaries}\label{sec:2}


\subsection{Geometric control theory for disturbance decoupling}\label{sec:2-1}

Let us consider the following {\it classical} linear time-invariant system:
\begin{align}
          \frac{dx(t)}{dt}&=Ax(t)+Bu(t), ~~ \nonumber\\
          y(t)&=Cx(t)+Du(t),
\label{eq_linearsystem}
\end{align}
where $x(t) \in \mathcal{X}:=\mathbb{R}^{n}$ is a vector of system variables, 
$u(t) \in \mathcal{U}:=\mathbb{R}^{m}$ and 
$y(t) \in \mathcal{Y}:=\mathbb{R}^{l}$ are vectors of input and output, 
respectively. 
$A, B, C$, and $D$ are real matrices. 
In the Laplace domain, the input-output relation is represented by 
\begin{align*}
         Y(s)=\Xi (s)U(s), ~~ \Xi (s)=C(sI-A)^{-1}B+D, 
\end{align*}
where $U(s)$ and $Y(s)$ are the Laplace transforms of $u(t)$ and $y(t)$, 
respectively. 
$\Xi(s)$ is called the {\it transfer function}. 
In this subsection, we assume $D=0$.

Now we describe the geometric control theory, for the disturbance 
decoupling problem \cite{Schumacher1980, Schumacher1982}. 
The following {\it invariant subspaces} play a key role in the theory. 

{\bf Definition 1:}~
Let $A : \mathcal{X} \rightarrow \mathcal{X}$ be a linear map. 
Then, a subspace $\mathcal{V} \subseteq \mathcal{X}$ is said 
to be $A$-{\it invariant}, if $A\mathcal{V} \subseteq \mathcal{V}$. 

{\bf Definition 2:}~
Given a linear map $A : \mathcal{X} \rightarrow \mathcal{X}$ 
and a subspace \Image $B$ $\subseteq \mathcal{X}$, a subspace 
$\mathcal{V} \subseteq \mathcal{X}$ is said to be $(A, B)$-{\it invariant}, 
if $A\mathcal{V} \subseteq \mathcal{V} \oplus \Image B$. 

{\bf Definition 3:}~ 
Given a linear map $A : \mathcal{X} \rightarrow \mathcal{X}$ 
and a subspace \Kernel $C$ $\subseteq \mathcal{X}$, a subspace 
$\mathcal{V}\subseteq \mathcal{X}$ is said to be $(C, A)$-{\it invariant}, 
if $A(\mathcal{V} \cap \Kernel C) \subseteq \mathcal{V}$. 

{\bf Definition 4:}~ 
Assume that $\mathcal{V}_{1}$ is $(C, A)$-invariant, $\mathcal{V}_{2}$ is 
$(A, B)$-invariant, and $\mathcal{V}_{1} \subseteq \mathcal{V}_{2}$. 
Then, $(\mathcal{V}_{1},\, \mathcal{V}_{2})$ is said to be a $(C,A,B)$-pair.

From Definitions 2 and 3, we have the following two lemmas.

{\bf Lemma 1:}~ 
$\mathcal{V} \subseteq \mathcal{X}$ is $(A, B)$-invariant if and only if 
there exists a matrix $F$ such that
$
             F \in 
             \mathcal{F}(\mathcal{V}) :=
                   \{ F : \mathcal{X} \rightarrow \mathcal{U}\,| \, 
                          (A+BF)\mathcal{V} \subseteq \mathcal{V} \}. 
$
%

{\bf Lemma 2:}~
$\mathcal{V} \subseteq \mathcal{X}$ is $(C, A)$-invariant if and only if 
there exists a matrix $G$ such that
$
              G \in 
                \mathcal{G}(\mathcal{V}) := 
                  \{ G : \mathcal{Y} \rightarrow \mathcal{X} \,| \, 
                         (A+GC)\mathcal{V} \subseteq \mathcal{V} \}. 
$
%

\begin{figure}[!t]
\begin{center}
\includegraphics[width=4cm,clip]{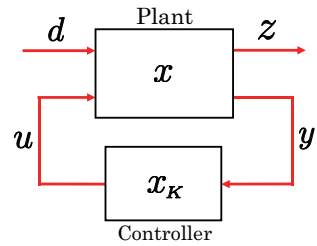}
\caption{General configuration for the disturbance decoupling via a dynamical feedback controller.}
\label{fig_feedbackframework}
\end{center}
\end{figure}

The disturbance decoupling problem is described as follows. 
The system of interest is represented, in an extended form of 
Eq.~\eqref{eq_linearsystem}, as 
%
\begin{align*}
         \frac{dx(t)}{dt}&=Ax(t)+Bu(t)+Ed(t), ~\\
         y(t)&=Cx(t), ~~
         z(t)=Hx(t), 
\end{align*}
%
where $d(t)$ is the disturbance and $z(t)$ is the output to be regulated. 
$E$ and $H$ are real matrices. 
The other output $y(t)$ may be used for constructing a feedback controller; 
see Fig.~\ref{fig_feedbackframework}. 
The disturbance $d(t)$ can degrade the control performance evaluated on $z(t)$. 
Thus it is desirable if we can modify the system structure by some means so 
that eventually $d(t)$ dose not affect at all on $z(t)$
\footnote{This condition is satisfied if the transfer function from $d(s)$ to $z(s)$ is 
zero for all $s$, for the modified system. Or equivalently, the controllable subspace
with respect to $d(t)$ is contained in the unobservable subspace with respect to $z(t)$.}.
This control goal is called the disturbance decoupling. 
Here we describe a specific feedback control method to achieve this goal; 
note that, as shown later, the direct-interaction method for linear quantum 
systems can also be described within this framework. 
The controller configuration is illustrated in Fig.~\ref{fig_feedbackframework}; 
that is, the system modification is carried out by combining an auxiliary 
system (controller) with the original system (plant), so that the whole 
closed-loop system satisfies the disturbance decoupling condition. 
The controller with variable 
${x}_{{\scriptscriptstyle K}} \in \mathcal{X}_{{\scriptscriptstyle K}}
:=\mathbb{R}^{n_{\scriptscriptstyle k}}$ is assumed to take the following form: 
%
\begin{align*}
\hspace{-0.1cm}
     \frac{dx_{{\scriptscriptstyle K}}(t)}{dt}
                   &=A_{\scriptscriptstyle K}x_{{\scriptscriptstyle K}}(t)
                       +B_{\scriptscriptstyle K}y(t), \, \\
      u(t)&=C_{\scriptscriptstyle K}x_{{\scriptscriptstyle K}}(t)
                       +D_{\scriptscriptstyle K}y(t), 
\end{align*}
%
where $A_{\scriptscriptstyle K} : \mathcal{X}_{{\scriptscriptstyle K}} \rightarrow 
\mathcal{X}_{{\scriptscriptstyle K}}$, 
$B_{\scriptscriptstyle K} : 
\mathcal{Y} \rightarrow \mathcal{X}_{{\scriptscriptstyle K}}$,
$C_{\scriptscriptstyle K} : 
\mathcal{X}_{{\scriptscriptstyle K}} \rightarrow \mathcal{U}$,  
and $D_{\scriptscriptstyle K} : \mathcal{Y} \rightarrow \mathcal{U}$ are real matrices. 
Then, the closed-loop system defined in the augmented space 
$\mathcal{X}_{{\scriptscriptstyle E}} := \mathcal{X} \oplus \mathcal{X}_{{\scriptscriptstyle K}}$ 
is given by 
\begin{align}
\label{eq_extended_system}
          \frac{d}{dt}\left[\begin{array}{c}
                x\\
                x_{{\scriptscriptstyle K}}\\
          \end{array}\right] 
         &=\left[\begin{array}{cc}
               A+BD_{\scriptscriptstyle K}C & BC_{\scriptscriptstyle K}\\
               B_{\scriptscriptstyle K}C & A_{\scriptscriptstyle K}\\
          \end{array}\right]
          \left[\begin{array}{c}
               x\\
               x_{{\scriptscriptstyle K}}\\
          \end{array}\right] 
        +\left[\begin{array}{c}
              E\\
              O\\
          \end{array}\right] d, ~ \nonumber \\
        z&=\left[\begin{array}{cc}
              H & O\\
         \end{array}\right]
        \left[\begin{array}{c}
              x\\
              x_{{\scriptscriptstyle K}}\\
         \end{array}\right].
\end{align}
The control goal is to design $(A_{\scriptscriptstyle K}, B_{\scriptscriptstyle K}, 
C_{\scriptscriptstyle K}, D_{\scriptscriptstyle K})$ so that, in 
Eq.~\eqref{eq_extended_system}, the disturbance signal $d(t)$ dose not 
appear in the output $z(t)$: see the below footnote.
Here, let us define 
\begin{align}
            A_{{\scriptscriptstyle E}}=\left[\begin{array}{cc}
                              A+BD_{\scriptscriptstyle K}C & BC_{\scriptscriptstyle K}\\
                              B_{\scriptscriptstyle K}C & A_{\scriptscriptstyle K}\\
                          \end{array}\right] , 
\label{eq_extended_matrixAe}
\end{align}
$\mathcal{B}= \Image B$, $\mathcal{C}= \Kernel C$, 
$\mathcal{E}= \Image E$, and $\mathcal{H}= \Kernel H$. 
Then, the following theorem gives the solvability condition for the disturbance 
decoupling problem.

{\bf Theorem 1:}~
For the closed-loop system \eqref{eq_extended_system}, the disturbance 
decoupling problem via the dynamical feedback controller has a solution 
if and only if there exists a $(C,A,B)$-pair $(\mathcal{V}_{1},\mathcal{V}_{2})$ 
satisfying 
\begin{align}
          \mathcal{E} \subseteq \mathcal{V}_{1} 
                    \subseteq \mathcal{V}_{2} \subseteq \mathcal{H}.
\label{eq_DDP_condition}
\end{align}

Note that this condition does not depend on the controller matrices to 
be designed. 
The following corollary can be used to check if the solvability condition 
is satisfied. 

{\bf Corollary 1:}~ 
For the closed-loop system \eqref{eq_extended_system}, the disturbance 
decoupling problem via the dynamical feedback controller has a solution 
if and only if
\begin{align}
           \mathcal{V}_{*}{\scriptscriptstyle (\mathcal{C}, \mathcal{E})}
              \subseteq \mathcal{V}^{*}{\scriptscriptstyle (\mathcal{B}, \mathcal{H})}, \nonumber
\end{align}
where $\mathcal{V}^{*}{\scriptscriptstyle (\mathcal{B}, \mathcal{H})}$ is 
the maximum element of $(A, B)$-invariant subspaces contained in $\mathcal{H}$, 
and $\mathcal{V}_{*}{\scriptscriptstyle (\mathcal{C}, \mathcal{E})}$ is 
the minimum element of $(C, A)$-invariant subspaces containing $\mathcal{E}$. 
These subspaces can be computed by the algorithms given in Appendix~A. 

Once the solvability condition described above is satisfied, then we can 
explicitly construct the controller matrices $(A_{\scriptscriptstyle K}, 
B_{\scriptscriptstyle K}, C_{\scriptscriptstyle K}, D_{\scriptscriptstyle K})$. 
The following intersection and projection subspaces play a key role for this purpose; 
that is, for a subspace $\mathcal{V}_{{\scriptscriptstyle E}} \subseteq 
\mathcal{X}_{{\scriptscriptstyle E}}= \mathcal{X} \oplus \mathcal{X}_{{\scriptscriptstyle K}}$, 
let us define
\begin{align*}
             \mathcal{V}_{{\scriptscriptstyle I}}:&= 
                        \left\{x \in \mathcal{X}~\Bigg| \left[\begin{array}{c}
                                    x\\
                                    O\\
                               \end{array}\right] 
                \in \mathcal{V}_{{\scriptscriptstyle E}} \right\}, ~ \\
             \mathcal{V}_{{\scriptscriptstyle P}}:&= 
                        \left\{x \in \mathcal{X}~\Bigg| \left[\begin{array}{c}
                                    x\\  
                                    x_{{\scriptscriptstyle K}}\\
                               \end{array}\right] 
                \in \mathcal{V}_{{\scriptscriptstyle E}},~ 
                \exists {x}_{{\scriptscriptstyle K}} \in 
                     \mathcal{X}_{{\scriptscriptstyle K}} \right\}.
\end{align*}
Then, the following theorem is obtained:

{\bf Theorem 2:}~ 
Suppose that $(\mathcal{V}_{1}, \mathcal{V}_{2})$ is a $(C, A, B)$-pair. 
Then, there exist $F \in \mathcal{F}(\mathcal{V}_{2})$, 
$G \in \mathcal{G}(\mathcal{V}_{1})$,
and $D_{\scriptscriptstyle K} : \mathcal{Y} \rightarrow \mathcal{U}$ 
such that $\Kernel F_{0} \supseteq \mathcal{V}_{1}$ and 
$\Image G_{0} \subseteq \mathcal{V}_{2}$ hold, where
$F_{0}=F-D_{\scriptscriptstyle K}C,~G_{0}=G-BD_{\scriptscriptstyle K}$.

Moreover, there exists $\mathcal{X}_{{\scriptscriptstyle K}}$ with 
${\rm dim}\,\mathcal{X}_{{\scriptscriptstyle K}}={\rm dim}\,
\mathcal{V}_{2} - {\rm dim}\,\mathcal{V}_{1}$, and $A_{{\scriptscriptstyle E}}$ 
has an invariant subspace $\mathcal{V}_{{\scriptscriptstyle E}} \subseteq 
\mathcal{X}_{{\scriptscriptstyle E}}$ such that $\mathcal{V}_{1}=\mathcal{V}_{{\scriptscriptstyle I}}$ 
and $\mathcal{V}_{2}=\mathcal{V}_{{\scriptscriptstyle P}}$. 
Also, $(A_{\scriptscriptstyle K}, B_{\scriptscriptstyle K}, C_{\scriptscriptstyle K})$ 
satisfies  
\begin{align}
\label{eq_DFC_characterize}
             &C_{\scriptscriptstyle K}N=F_{0}, ~~ \nonumber \\
             &B_{\scriptscriptstyle K}=-NG_{0}, ~~ \nonumber \\
             &A_{\scriptscriptstyle K}N=N(A+BF_{0}+GC),
\end{align}
where $N : \mathcal{V}_{2} \rightarrow \mathcal{X}_{{\scriptscriptstyle K}}$ 
is a linear map satisfying $\Kernel N=\mathcal{V}_{1}$.

In fact, under the condition given in Theorem 2, let us define the following 
augmented subspace $\mathcal{V}_{{\scriptscriptstyle E}} \subseteq \mathcal{X}_{{\scriptscriptstyle E}}$:
\begin{align*}
       \mathcal{V}_{{\scriptscriptstyle E}} :=& \left\{
           \left[\begin{array}{c}
                        x\\
                        Nx\\
           \end{array}\right]  
       \Bigg|~ x \in \mathcal{V}_{2} \right\}.
\end{align*} 
Then, $\mathcal{V}_{1}=\mathcal{V}_{{\scriptscriptstyle I}}$ and 
$\mathcal{V}_{2}=\mathcal{V}_{{\scriptscriptstyle P}}$ 
hold, and we have 
\begin{align*}
     A_{{\scriptscriptstyle E}}\left[\begin{array}{c}
                   x\\
                   Nx\\
          \end{array}\right]  
     &=\left[\begin{array}{cc}
                   A+BD_{\scriptscriptstyle K}C & BC_{\scriptscriptstyle K}\\
                   B_{\scriptscriptstyle K}C & A_{\scriptscriptstyle K}\\
          \end{array}\right]
          \left[\begin{array}{c}
                   x\\
                   Nx\\
          \end{array}\right]  \nonumber\\
     &=\left[\begin{array}{c}
                   (A+BF)x\\
                   N(A+BF)x\\
          \end{array}\right] 
  \in \mathcal{V}_{{\scriptscriptstyle E}},
\end{align*} 
implying that $\mathcal{V}_{{\scriptscriptstyle E}}$ is actually $A_{{\scriptscriptstyle E}}$-invariant. 
Now suppose that Theorem~1 holds, and let us take the $(C,A,B)$-pair 
$(\mathcal{V}_{1}, \mathcal{V}_{2})$ satisfying Eq.~\eqref{eq_DDP_condition}. 
Then, together with the above result 
($A_{{\scriptscriptstyle E}}\mathcal{V}_{{\scriptscriptstyle E}} \subseteq \mathcal{V}_{{\scriptscriptstyle E}}$), 
we have $\Image [E^{\top}~O]^{\top} \subseteq \mathcal{V}_{{\scriptscriptstyle E}} 
\subseteq \Kernel [H~O]$. 
This implies that $d(t)$ must be contained in the unobservable subspace 
with respect to $z(t)$, and thus the disturbance decoupling is realized.


\subsection{Linear quantum systems}\label{sec:2-2}

Here we describe a general linear quantum system composed of $n$ bosonic 
subsystems. 
The $j$-th mode can be modeled as a harmonic oscillator with the canonical 
conjugate pairs (or quadratures) $\hat{q}_{j}$ and $\hat{p}_{j}$ satisfying the 
canonical commutation relation (CCR) 
$\hat{q}_{j}\hat{p}_{k}-\hat{p}_{k}\hat{q}_{j}=i \delta_{jk}$. 
Let us define the vector of quadratures as 
$\hat{x}=[\hat{q}_{1}, \hat{p}_{1}, \ldots, \hat{q}_{n}, \hat{p}_{n}]^{\top}$. 
Then, the CCRs are summarized as 
\begin{align*} 
       &\hat{x}\hat{x}^{\top}-(\hat{x}\hat{x}^{\top})^{\top} 
                = i \Sigma_{n}, ~~\\
       &\Sigma_{n} 
                = {\rm diag}\{\Sigma, \ldots, \Sigma\},~~~
        \Sigma=\left[\begin{array}{cc}
                            0 & 1 \\
                          -1 & 0 \\
                       \end{array}\right]. 
\end{align*}
Note that $\Sigma_{n}$ is a $2n \times 2n$ block diagonal matrix. 
The linear quantum system is an open system coupled to $m$ environment 
fields via the interaction Hamiltonian 
$\hat{H}_{\rm int}=i\sum_{j=1}^m (\hat{L}_j \hat{A}_j^*-\hat{L}_j^* \hat{A}_j)$, 
where $\hat{A}_j(t)$ is the field annihilation operator satisfying 
$\hat{A}_j(t)\hat{A}_k^{*}(t')-\hat{A}_k^{*}(t')\hat{A}_j(t)=\delta_{jk}\delta(t-t')$. 
Also $\hat{L}_j$ is given by $\hat{L}_j=c_j^\top\hat{x}$ with 
$c_j\in{\mathbb C}^{2n}$. 
In addition, the system is driven by the Hamiltonian 
$\hat{H}=\hat{x}^\top R\hat{x}/2$ with 
$R=R^\top \in \mathbb{R}^{2n \times 2n}$. 
Then, the Heisenberg equation of $\hat{x}$ is given by
\begin{align}
             \frac{d\hat{x}(t)}{dt}=A\hat{x}(t)+\sum_{j=1}^m B_j\hat{W}_j(t), 
\label{eq_LQS_dynamics}
\end{align}
where $\hat{W}_j(t)$ is defined by 
\[
        \hat{W}_j
        = \left[\begin{array}{cc}
             \hat{Q}_{j} \\
             \hat{P}_{j} \\
           \end{array}\right]
       = \left[\begin{array}{cc}
             (\hat{A}_j+\hat{A}_j^*)/\sqrt{2} \\
             (\hat{A}_j-\hat{A}_j^*)/\sqrt{2} i \\
           \end{array}\right].
\]
The matrices are given by 
$A=\Sigma_n(R + \sum_{j=1}^m C_j^\top\Sigma C_j/2)$ and 
$B_j=\Sigma_n C_j^\top\Sigma$ 
with $C_j=\sqrt{2}[\Re(c_j), \Im(c_j)]^\top \in \mathbb{R}^{2 \times 2n}$. 
Also, the instantaneous change of the field operator $\hat{W}_j(t)$ via the 
system-field coupling is given by 
\begin{align}
           \hat{W}_j^{\rm out}(t)=C_j\hat{x}(t)+\hat{W}_j(t). 
\label{eq_LQS_output}
\end{align}
Summarizing, the linear quantum system is characterized by the dynamics 
\eqref{eq_LQS_dynamics} and the output \eqref{eq_LQS_output}, which are 
exactly of the same form as those in Eq.~\eqref{eq_linearsystem} ($l=m$ in this case). 
However note that the system matrices have to satisfy the above-described 
special structure, which is equivalently converted to the following 
{\it physical realizability condition} \cite{James2008}: 
\begin{equation}
\label{phys real condition}
        A\Sigma_n+\Sigma_n A^{\top}
                 + \sum_{j=1}^m B_j \Sigma B_j^{\top}=O,~
        B_j=\Sigma_n C_j^{\top}\Sigma. 
\end{equation}
%


\subsection{Weak signal sensing, SQL, and BAE}\label{sec:2-3}

\begin{figure}[!t]
 \begin{center}
 \includegraphics[width=5.2cm,clip]{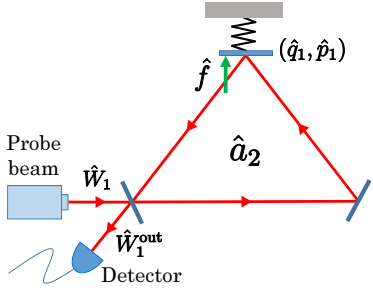}
 \end{center} 
 \caption{Opto-mechanical system for weak signal sensing.}
 \label{fig_opticalsensor}
\end{figure}

The opto-mechanical oscillator illustrated in Fig.~\ref{fig_opticalsensor} is a 
linear quantum system, which serves as a sensor for a very weak signal. 
Let $\hat{q}_{1}$ and $\hat{p}_{1}$ be the oscillator's position and momentum 
operators, and $\hat{a}_{2}=(\hat{q}_{2}+i \hat{p}_{2})/\sqrt{2}$ represents 
the annihilation operator of the cavity mode. 
The system Hamiltonian is given by 
$\hat{H}=\omega_{\scriptscriptstyle m}(\hat{q}_1^2+\hat{p}_1^2)/2
-g\hat{q}_1\hat{q}_2$; that is, the oscillator's free evolution with resonant 
frequency $\omega_{\scriptscriptstyle m}$ plus the linearized radiation 
pressure interaction between the oscillator and the cavity field with coupling 
strength $g$. 
The system couples to an external probe field (thus $m=1$) via the coupling 
operator $\hat{L}_1=\sqrt{\kappa}\hat{a}_2$, with $\kappa$ the coupling 
constant between the cavity and probe fields. 
The corresponding matrix $R$ and vector $c_1$ are then given by 
\begin{align*}
           R=\left[\begin{array}{cccc}
                   \omega_{\scriptscriptstyle m} &0&-g&0\\
                   0&\omega_{\scriptscriptstyle m} &0&0\\
                  -g&0&0&0\\
                  0&0&0&0\\
                \end{array}\right],~~
           c_{1}=\sqrt{\frac{\kappa}{2}}
                \left[\begin{array}{c}
                   0 \\ 0 \\ 1 \\ i \\
                \end{array}\right].
\end{align*}
The oscillator is driven by an unknown force $\hat{f}(t)$ with coupling constant $\gamma$ ; 
then the vector of system variables $\hat{x}=[\hat{q}_1, \hat{p}_1, \hat{q}_2, \hat{p}_2]^\top$ satisfies 
\begin{equation*}
         \frac{d\hat{x}}{dt}
                   =A\hat{x}+B_1\hat{W}_1+b \hat{f},~~
          \hat{W}_{1}^{\rm out}
                   =C_1\hat{x}+\hat{W}_1,
\end{equation*}
where
\begin{align}
           A&=\left[\begin{array}{cccc}
                      0&\omega_{\scriptscriptstyle m}&0&0\\
                      -\omega_{\scriptscriptstyle m} &0 &g&0\\
                      0&0&-\kappa/2 &0\\
                      g&0&0&-\kappa/2 \\
                  \end{array}\right], \,
           b=\sqrt{\gamma}\left[\begin{array}{c}
                      0 \\
                      1 \\
                      0 \\
                      0 \\
               \end{array}\right], \nonumber \\
          C_1&=-B_1^{\top}
              =\left[\begin{array}{cccc}
                     0 & 0 & \sqrt{\kappa} & 0\\
                     0 & 0 & 0 & \sqrt{\kappa}\\
                \end{array}\right], \nonumber \\
          \hat{W}_{1}
                 &=[\hat{Q}_{1}, \hat{P}_{1}]^{\top}, ~~
         \hat{W}_{1}^{\rm out}
                  =[\hat{Q}_{1}^{\rm out}, \hat{P}_{1}^{\rm out}]^{\top}.
\label{eq_matrix_originalplant}
\end{align}
Note that we are in the rotating frame at the frequency of the probe field. 
These equations indicate that the information about $\hat{f}$ can be extracted 
by measuring $\hat{P}_1^{\rm out}$ by a homodyne detector. 
Actually the measurement output in the Laplace domain is given by
\begin{align}
          \hat{P}_{1}^{\rm out}(s)
                =\Xi_{f}(s) \hat{f}(s)+\Xi_{Q}(s) \hat{Q}_{1}(s)
                                             +\Xi_{P}(s) \hat{P}_{1}(s),
\label{eq_measurement_output}
\end{align}
where $\Xi_{f}$, $\Xi_{Q}$, and $\Xi_{P}$ are transfer functions given by 
\begin{align*}
\hspace{-0.8cm}
         \Xi_{f}(s)&=\frac{g\omega_{\scriptscriptstyle m}\sqrt{\gamma\kappa}}
                                       {(s^2+\omega_{\scriptscriptstyle m}^2)(s+\kappa/2)},  ~~\\
         \Xi_{Q}(s)&=-\frac{g^2 \omega_{\scriptscriptstyle m}\kappa}
                                         {(s^2+\omega_{\scriptscriptstyle m}^2)(s+\kappa/2)^2}, ~~
         \Xi_{P}(s)=\frac{s-\kappa/2}{s+\kappa/2}. 
\end{align*}
Thus, $\hat{P}_1^{\rm out}$ certainly contains $\hat{f}$. 
Note however that it is subjected to two noises. 
The first one, $\hat{Q}_{1}$, is the back-action noise, which is due to the 
interaction between the oscillator and the cavity. 
The second one, $\hat{P}_{1}$, is the shot noise, which inevitably appears. 
Now, the normalized output is given by
\[
             y_{1}(s)
                    =\frac{\hat{P}_{1}^{\rm out}(s)}{\Xi_{f}(s)}
                    =\hat{f}(s) + \frac{\Xi_{Q}(s)}{\Xi_{f}(s)}\hat{Q}_{1}(s)
                              + \frac{\Xi_{P}(s)}{\Xi_{f}(s)}\hat{P}_{1}(s),
\]
and the normalized noise power spectral density of $y_{1}$ in the Fourier domain 
$(s=i\omega)$ is calculated as follows:
\begin{align*}
        &S(\omega) =\langle |y_{1}-\hat{f}|^2 \rangle
                          =\left| \frac{\Xi_{Q}}{\Xi_{f}} \right| ^2 
                                 \langle|\hat{Q}_1|^2 \rangle 
                           + \left| \frac{\Xi_{P}}{\Xi_{f}} \right| ^2 
                                 \langle |\hat{P}_1|^2 \rangle  \\
                           &\geq 2\sqrt{\frac{|\Xi_{Q}|^2|\Xi_{P}|^2}{|\Xi_{f}|^4} 
                                 \langle |\hat{Q}_1|^2 \rangle 
                                 \langle |\hat{P}_1|^2 \rangle }
                          \geq\frac{| \omega^2-\omega_{\scriptscriptstyle m}^2 | }
                                              {\gamma\omega_{\scriptscriptstyle m}}
                           =S_{\scriptscriptstyle \rm SQL}(\omega).
\end{align*}
The lower bound is called the SQL. 
Note that the last inequality is due to the Heisenberg uncertainty relation of the normalized noise power, i.e.,
$\langle|\hat{Q}_1|^2 \rangle \langle|\hat{P}_1|^2 \rangle \geq 1/4$. 
Hence, the essential reason why SQL appears is that $\hat{P}_{1}^{\rm out}$ 
contains both the back-action noise $\hat{Q}_1$ and the shot noise $\hat{P}_1$. 
Therefore, toward the high-precision detection of $\hat{f}$, we need BAE; 
that is, the system structure should be modified by some means so that the 
back-action noise is completely evaded in the output signal (note that the shot noise can never be evaded). 
The condition for BAE can be expressed in terms of the transfer function 
as follows \cite{Tsang2010, YamamotoCFvsMF2014}; 
i.e., for the modified (controlled) sensor, the transfer function from 
the back-action noise to the measurement output must satisfy 
\begin{align}
              \Xi_{Q}(s)=0, ~~~\forall s
\label{eq_perfectBAE1}.
\end{align}
Equivalently, $\hat{P}_{1}^{\rm out}$ contains only the shot noise $\hat{P}_{1}$; 
hence, in this case the signal to noise ratio can be further improved by 
injecting a $\hat{P}_{1}$-squeezed (meaning $\langle|\hat{P}_{1}|^2 \rangle < 1/2$) 
probe field into the system.


\section{Coherent feedback control for back-action evasion}\label{sec:3}


\subsection{Coherent and measurement-based feedback control}\label{sec:3-1}

\begin{figure}[!t]
\begin{center}
\includegraphics[width=7.7cm,clip]{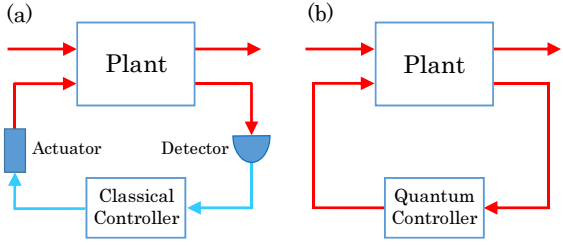}
\caption{General configurations of feedback control for a given plant quantum system: 
(a) measurement-based feedback and (b) coherent feedback.}
\label{fig_MFvsCF}
\end{center}
\end{figure}

There are two schemes for controlling a quantum system via feedback. 
The first one is the {\it measurement-based feedback} 
\cite{Belavkin1999, Bouten2009, WisemanBook, KurtBook} illustrated in 
Fig.~\ref{fig_MFvsCF}~(a). 
In this scheme, we measure the output fields and feed the measurement 
results back to control the plant system. 
On the other hand, in the {\it coherent feedback} scheme 
\cite{James2008,Wiseman1994, Yanagisawa-2003,Mabuchi2008, Iida2012} 
shown in Fig.~\ref{fig_MFvsCF}~(b), the feedback loop dose not contain 
any measurement component and the plant system is controlled by another quantum system. 
Recently we find several works comparing the performance of these two schemes 
\cite{Wiseman1994, NurdinLQG2009, Hamerly2012, Jacobs2014, Devoret2016}. 
In particular, it was shown in \cite{YamamotoCFvsMF2014} that there are 
some control tasks that cannot be achieved by any measurement-based feedback 
but can be done by a coherent one. 
More specifically, those tasks are realizing BAE measurement, generating 
a quantum non-demolished variable, and generating a decoherence-free 
subsystem; 
in our case, of course, the first one is crucial. 
Hence, here we aim to develop a theory for designing a coherent feedback 
controller such that the whole controlled system accomplishes BAE.


\subsection{Coherent feedback for BAE}\label{sec:3-2}

As discussed in Section \ref{sec:2-1}, the geometric control theory for 
disturbance decoupling problem is formulated for the controlled system 
with special structure \eqref{eq_extended_system}; 
in particular, the coefficient matrix of the disturbance $d(t)$ is of the form $[E^\top, O]^\top$ 
and that of the state vector in the output $z(t)$ is $[H, O]$. 
Here we consider a class of coherent feedback configuration such that 
the whole closed-loop system dynamics has this structure, in order for 
the geometric control theory to be directly applicable.

\begin{figure}[!t]
\begin{center}
\includegraphics[width=5cm,clip]{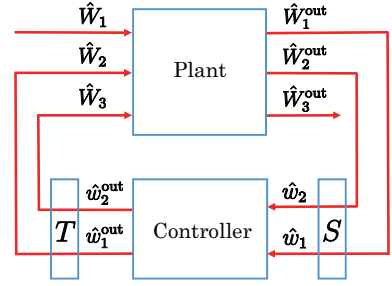}
\caption{Coherent feedback control of the 3 input-output plant system via the 2 input-output controller.}
\label{fig_3I3O_general}
\end{center}
\end{figure}

First, for the plant system given by Eqs.~\eqref{eq_LQS_dynamics} and 
\eqref{eq_LQS_output}, we assume that the system couples to all the probe fields  
in the same way; i.e., 
\begin{align}
       B_{j}=B~~~\forall j. 
\label{eq_plant_symmetry}
\end{align}
This immediately leads to $C_{j}=C~\forall j$. 
Next, as the controller, we take the following special linear quantum system 
with $(m-1)$ input-output fields: 
\begin{align}
\label{eq_dynamical_controller} 
          \frac{d\hat{x}_{\scriptscriptstyle K}}{dt}
                 &=A_{\scriptscriptstyle K} \hat{x}_{\scriptscriptstyle K}
                     +\sum_{j=1}^{m-1} B_{{\scriptscriptstyle K}} \hat{w}_{j},  \nonumber\\
         \hat{w}_{j}^{\rm out}
                 &=C_{{\scriptscriptstyle K}} \hat{x}_{\scriptscriptstyle K}
                               +\hat{w}_{j}~~~(j=1,\,2,\, \ldots, m-1),
\end{align}
where the matrices $(A_{\scriptscriptstyle K}, B_{\scriptscriptstyle K}, 
C_{\scriptscriptstyle K})$ satisfy the physical realizability condition 
\eqref{phys real condition}. 
Note that, corresponding to the plant structure, we assumed that the controller 
couples to all the fields in the same way, specified by $C_{\scriptscriptstyle K}$. 
Here we emphasize that the number of channels, $m$, should be as small as 
possible from a viewpoint of implementation; 
hence in this paper let us consider the case $m=3$. 
Now, we consider the coherent feedback connection illustrated in 
Fig.~\ref{fig_3I3O_general}, i.e., 
\begin{align*}
               &\hat{w}_{1}=S_{1}\hat{W}_{1}^{\rm out},~~ 
                   \hat{w}_{2}=S_{2}\hat{W}_{2}^{\rm out}, ~~ \\
                &\hat{W}_{2}=T_1\hat{w}_{1}^{\rm out},~~ 
                   \hat{W}_{3}=T_{2}\hat{w}_{2}^{\rm out},
\end{align*}
where $S_{j}$ and $T_{j}$ are $2 \times 2$ unitary matrices representing 
the scattering process of the fields; 
recall that the scattering process $\hat{A}^{\rm out}=e^{i\theta}\hat{A}$ 
with $\theta \in \mathbb{R}$ the phase shift can be represented in the 
quadrature form as 
\begin{equation*}
           \left[\begin{array}{cc}
             \hat{Q}^{\rm out} \\
             \hat{P}^{\rm out} \\
           \end{array}\right]
           = S(\theta)
              \left[\begin{array}{cc}
                 \hat{Q} \\
                 \hat{P} \\
              \end{array}\right]
            =\left[\begin{array}{cc}
                   \cos \theta &-\sin \theta \\
                    \sin \theta &\cos \theta \\
               \end{array}\right]
               \left[\begin{array}{cc}
                 \hat{Q} \\
                 \hat{P} \\
              \end{array}\right].
\label{eq_phaseshifter}
\end{equation*}

Combining the above equations, we find that the whole closed-loop system 
with the augmented variable 
$\hat{x}_{{\scriptscriptstyle E}}=[\hat{x}^{\top},\hat{x}_{\scriptscriptstyle K}^{\top}]^{\top}$ 
is given by
\begin{align}
\label{eq_3I3Oclosed2}
             \frac{d\hat{x}_{{\scriptscriptstyle E}}}{dt}
                                  &=A_{{\scriptscriptstyle E}}\hat{x}_{{\scriptscriptstyle E}}
                                              +B_{{\scriptscriptstyle E}}\hat{W}_{1}+b_{{\scriptscriptstyle E}}\hat{f}, \nonumber\\
             \hat{W}_{3}^{\rm out}
                                  &=C_{{\scriptscriptstyle E}}\hat{x}_{{\scriptscriptstyle E}}
                                              +D_{{\scriptscriptstyle E}}\hat{W}_{1},
\end{align}
where
%
%
\begin{align*}
A_{{\scriptscriptstyle E}}
&=\left[\begin{array}{c}
A+B\{ T_{1}S_{1}+T_{2}S_{2}(T_{1}S_{1}+I_{2})\} C  \\ [1.0ex]
B_{\scriptscriptstyle K}\{ (I_{2}+S_{2}T_{1})S_{1}+S_{2}\} C 
\end{array}\right. \\
&~~~~~~~~~~~~~~~~~~~~~~~~~~~\left. \begin{array}{c}
 B\{ T_{1}+T_{2}(I_{2}+S_{2}T_{1})\} C_{\scriptscriptstyle K} \\ [1.0ex]
 A_{\scriptscriptstyle K}+B_{\scriptscriptstyle K}S_{2}T_{1}C_{\scriptscriptstyle K}
\end{array}\right]. \\
B_{{\scriptscriptstyle E}}
&=\left[\begin{array}{c}
                            B(I_{2}+T_{1}S_{1}+T_{2}S_{2}T_{1}S_{1}) \\ [1.0ex]
                            B_{{\scriptscriptstyle K}}(I_{2}+S_{2}T_{1})S_{1} \\
                        \end{array}\right] \\
C_{{\scriptscriptstyle E}}
&=\left[\begin{array}{cc}
                           (T_{2}S_{2}T_{1}S_{1}+T_{2}S_{2}+I_{2})C ~
                           & ~T_{2}(S_{2}T_{1}+I_{2})C_{\scriptscriptstyle K}\\
                        \end{array}\right] \\
D_{{\scriptscriptstyle E}}&=T_{2}S_{2}T_{1}S_{1}, ~~
b_{{\scriptscriptstyle E}}=\left[\begin{array}{cc}
                           b^{\top} & O\\
                       \end{array}\right]^{\top}.
\end{align*}
Therefore, the desired system structure of the form \eqref{eq_extended_system} 
is realized if we take 
\begin{align}
       S_{2}T_{1}=-I_{2}.
\label{eq_special_structure1}
\end{align}
In addition, it is required that the back-action noise $\hat{Q}_{1}$ 
dose not appear directly in $\hat{P}_{3}^{\rm out}$, which can be realized by 
taking 
\begin{align}
        D_{{\scriptscriptstyle E}}=-T_{2}S_{1}=\pm I_{2}.
\label{eq_special_structure2}
\end{align}
Here we set $S_j$ and $T_j$ to be the $\pi/2$-phase shifter 
(see Fig.~\ref{fig_3I3OBAE}) to satisfy the above conditions \eqref{eq_special_structure1} and 
\eqref{eq_special_structure2};
\begin{align}
\label{eq_phaseshifter_matching}
          S_{j}=T_{j}=S
          =\left[\begin{array}{cc}
             0 & -1 \\
             1 & 0  \\
           \end{array}\right] ~~(j=1,2). 
\end{align}
As a consequence, we end up with 
\begin{align}
     &A_{{\scriptscriptstyle E}}=\left[\begin{array}{cc}
                       A-BC & BSC_{\scriptscriptstyle K} \\ 
                       B_{\scriptscriptstyle K}SC & 
                       A_{\scriptscriptstyle K}-B_{\scriptscriptstyle K}C_{\scriptscriptstyle K} 
                    \end{array}\right], ~~
       B_{{\scriptscriptstyle E}}=\left[\begin{array}{c}
                       B\\
                       O \\
                    \end{array}\right], 
\nonumber \\
     &
      C_{{\scriptscriptstyle E}}=\left[\begin{array}{cc}
                    C & O \\
                \end{array}\right], ~~
      D_{{\scriptscriptstyle E}}=I_{2}, ~~
      b_{{\scriptscriptstyle E}}=\left[\begin{array}{cc}
                    b^{\top} & O \\
                \end{array}\right]^{\top}.
\label{extended_matrices}
\end{align}
This is certainly of the form \eqref{eq_extended_system} with 
$D_{\scriptscriptstyle K}=-I_2$. 
Hence, we can now directly apply the geometric control theory to design 
a coherent feedback controller achieving BAE; 
that is, our aim is to find $(A_{\scriptscriptstyle K}, B_{\scriptscriptstyle K}, 
C_{\scriptscriptstyle K})$ such that, for the closed-loop system 
\eqref{eq_3I3Oclosed2}, the back-action noise $\hat{Q}_{1}$ (the first element 
of $\hat W_1$) does not appear in the measurement output $\hat P_3^{\rm out}$ 
(the second element of $\hat W_3^{\rm out}$). 
Note that those matrices must satisfy the physical 
realizability condition \eqref{phys real condition}, and thus they cannot be 
freely chosen. 
We need to take into account this additional constraint when applying 
the geometric control theory to determine the controller matrices.


\subsection{Coherent feedback realization of BAE in the opto-mechanical system}\label{sec:3-3}

\begin{figure}[!t]
\begin{center}
\includegraphics[width=6.5cm,clip]{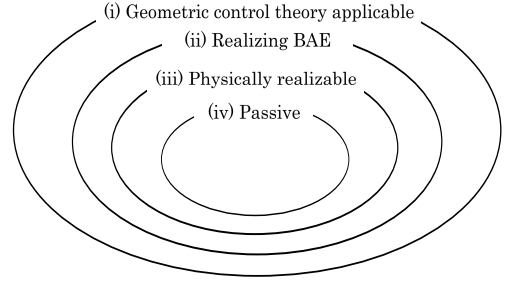}
\caption{
The set of controllers satisfying the condition in each step. For the controller to be a quantum system, 
it must be included in the set (iii). In the set (iv), all the controllers are equivalent up to the phase shift. 
}
\label{fig_controllerconstraint}
\end{center}
\end{figure}

Here we apply the coherent feedback scheme elaborated in Section~\ref{sec:3-2} to 
the opto-mechanical system studied in Section~\ref{sec:2-3}. 
The goal is, as mentioned before, to determine the controller matrices 
$(A_{\scriptscriptstyle K}, B_{\scriptscriptstyle K}, C_{\scriptscriptstyle K})$ 
such that the closed-loop system achieves BAE. 
Here, we provide a step-by-step procedure to solve this problem; 
the relationships of the class of controllers determined in each step is depicted 
in Fig.~\ref{fig_controllerconstraint}. 

\begin{figure}[!t]
\begin{center}
\includegraphics[width=7cm,clip]{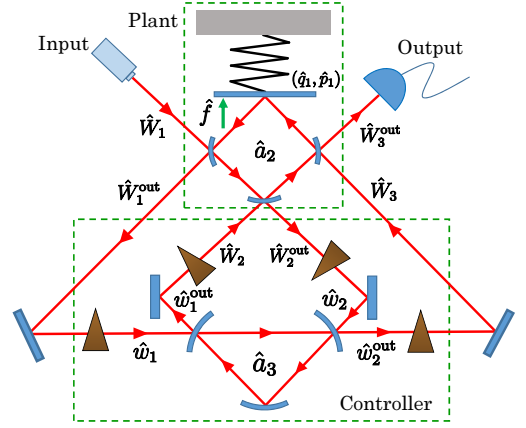}
\caption{
Coherent feedback controlled system composed of the opto-mechanical oscillator, for realizing BAE. 
The triangle represents the $\pi/2$-phase shifter corresponding to Eq.~\eqref{eq_phaseshifter_matching}. 
}
\label{fig_3I3OBAE}
\end{center}
\end{figure}

{\bf (i)} 
First, to apply the geometric control theory developed above, we need to modify 
the plant system so that it is a 3 input-output linear quantum system; 
here we consider the plant composed of a mechanical oscillator and a 3-ports optical cavity, 
shown in Fig.~\ref{fig_3I3OBAE}. 
As assumed before, those ports have the same coupling constant $\kappa$. 
In this case the matrix $A$ given in Eq.~\eqref{eq_matrix_originalplant} is 
replaced by 
\begin{align}
           A=\left[\begin{array}{cccc}
                     0&\omega_{\scriptscriptstyle m} &0&0\\
                     -\omega_{\scriptscriptstyle m}& 0&g &0\\
                     0&0&-3\kappa/2 &0\\
                     g &0&0&-3\kappa/2 \\
                \end{array}\right] .
\nonumber 
\end{align} 
Now we focus only on the back-action noise $\hat{Q}_{1}$ and the measurement output 
$\hat{P}_{3}^{\rm out}$; hence the closed-loop system \eqref{eq_3I3Oclosed2} and
\eqref{extended_matrices}, which ignores the shot noise term in the dynamical 
equation, is given by 
\begin{align*}
\hspace{-0.8cm}
          \frac{d\hat{x}_{{\scriptscriptstyle E}}}{dt}
              &=\left[\begin{array}{cc}
                       A-BC & BSC_{\scriptscriptstyle K} \\ 
                       B_{\scriptscriptstyle K}SC & 
                        A_{\scriptscriptstyle K}
                              -B_{\scriptscriptstyle K}C_{\scriptscriptstyle K} 
                    \end{array}\right]
                      \hat{x}_{{\scriptscriptstyle E}}  \\
&~~~~~~~~~~~~~~~~~~~~~~~~~~~~~~ + \left[\begin{array}{c}
                        E \\
                        O \\
                    \end{array}\right] \hat{Q}_{1}
                + \left[\begin{array}{c}
                        b\\
                        O\\
                   \end{array}\right] \hat{f},  \\
            \hat{P}_{3}^{\rm out}
                 &=\left[\begin{array}{cc}
                         H & O \\
                      \end{array}\right] \hat{x}_{{\scriptscriptstyle E}}
                    +\hat{P}_{1}, 
\end{align*}
where $B=B_1$, $C=C_1$, and $b$ are given in 
Eq.~\eqref{eq_matrix_originalplant}, and 
\begin{align*}
      E=-\sqrt{\kappa}\left[\begin{array}{cccc}
                        0&0&1&0 \\
                    \end{array}\right]^{\top},~~~ 
      H=\sqrt{\kappa}\left[\begin{array}{cccc}
                        0&0&0&1 \\
                    \end{array}\right].
\end{align*}
This system is certainly of the form \eqref{eq_extended_system}, where now $D_{\scriptscriptstyle K}=-I_2$.

{\bf (ii)} 
In the next step we apply Theorem~1 to check if there exists a feedback 
controller such that the above closed-loop system achieves BAE; 
recall that the necessary and sufficient condition is Eq.~\eqref{eq_DDP_condition}, 
i.e., $\mathcal{E} \subseteq \mathcal{V}_{1} \subseteq \mathcal{V}_{2} 
\subseteq \mathcal{H}$, where now 
\begin{align*}
       \mathcal{E}&=\Image E={\rm span} \left\{
                    \left[\begin{array}{c}
                         0\\
                         0\\
                         1\\
                         0\\
                    \end{array}\right] \right\}, \\
       \mathcal{H}&=\Kernel H={\rm span} \left\{
                    \left[\begin{array}{c}
                         1\\
                         0\\
                         0\\
                         0\\
                    \end{array}\right] ,
                    \left[\begin{array}{c}
                         0\\
                         1\\
                         0\\
                         0\\
                    \end{array}\right] ,
                    \left[\begin{array}{c}
                         0\\
                         0\\
                         1\\
                         0\\
                    \end{array}\right] \right\}. 
\end{align*}
To check if this solvability condition is satisfied, we use Corollary 1; 
from
$\mathcal{E} \cap \mathcal{C}=\Image E \cap \Kernel C =\phi$ and 
$\mathcal{H} \oplus \mathcal{B} = \Kernel H \oplus \Image B =\mathbb{R}^{4}$,
the algorithms given in Appendix~A yield 
\begin{align}
       \mathcal{V}_{*}{\scriptscriptstyle (\mathcal{C}, \mathcal{E})}
                    =\mathcal{E},~~~
       \mathcal{V}^{*}{\scriptscriptstyle (\mathcal{B}, \mathcal{H})}
                    =\mathcal{H}, 
\label{eq_CABpairs}
\end{align}
implying that the condition in Corollary 1, i.e., 
$\mathcal{V}_{*}{\scriptscriptstyle (\mathcal{C}, \mathcal{E})} \subseteq 
\mathcal{V}^{*}{\scriptscriptstyle (\mathcal{B}, \mathcal{H})}$, is satisfied. 
Thus, we now see that the BAE problem is solvable, as long as there is no 
constraint on the controller parameters.

Next we aim to determine the controller matrices 
$(A_{\scriptscriptstyle K}, B_{\scriptscriptstyle K}, C_{\scriptscriptstyle K})$, 
using Theorem~2. First we set 
$\mathcal{V}_{1}=\mathcal{V}_{*}{\scriptscriptstyle (\mathcal{C}, \mathcal{E})} = \mathcal{E}$ and 
$\mathcal{V}_{2}=\mathcal{V}^{*}{\scriptscriptstyle (\mathcal{B}, \mathcal{H})}= \mathcal{H}$;
note that $(\mathcal{V}_{1}, \mathcal{V}_{2})$ is a $(C, A, B)$-pair.
Then, from Theorem 2, there exists a feedback controller with dimension
${\rm dim}\,\mathcal{X}_{\scriptscriptstyle K}={\rm dim}\,\mathcal{V}_{2}-{\rm dim}\,\mathcal{V}_{1}=2$. 
Moreover, noting again that $D_{\scriptscriptstyle K}=-I_2$, there exist matrices 
$F \in \mathcal{F}(\mathcal{V}_{2})$, $G \in \mathcal{G}(\mathcal{V}_{1})$, 
and $N$ 
such that
\begin{align*}
     \Kernel F_{0}&=\Kernel (F+C) \supseteq \mathcal{V}_{1}, \\
     \Image G_{0}&=\Image (G+B) \subseteq \mathcal{V}_{2}, ~~
     \Kernel N=\mathcal{V}_{1}. 
\end{align*}
These conditions lead to
\begin{align*}
\hspace{-0.7cm}
              F&=\left[\begin{array}{cccc}
                        f_{\scriptscriptstyle 11} & f_{\scriptscriptstyle 12} 
                            & -\sqrt{\kappa} & f_{\scriptscriptstyle 14} \\
                        \frac{g}{\sqrt{\kappa}}&0&0&f_{\scriptscriptstyle 24} \\
                     \end{array}\right], ~~
              G=\left[\begin{array}{cc}
                        0&g_{\scriptscriptstyle 12}\\
                        -\frac{g}{\sqrt{\kappa}}&g_{\scriptscriptstyle 22}\\
                        g_{\scriptscriptstyle 31}&g_{\scriptscriptstyle 32}\\
                        0&\sqrt{\kappa}\\
                     \end{array}\right],  \\
              N&=\left[\begin{array}{cccc}
                        n_{\scriptscriptstyle 11} & n_{\scriptscriptstyle 12} 
                             & 0 & n_{\scriptscriptstyle 14} \\
                        n_{\scriptscriptstyle 21} & n_{\scriptscriptstyle 22}
                             & 0 & n_{\scriptscriptstyle 24}\\
                     \end{array}\right], 
\end{align*}
where $f_{ij}, g_{ij}$, and $ n_{ij}$  are free parameters. Then the controller matrices 
$(A_{\scriptscriptstyle K}, B_{\scriptscriptstyle K}, C_{\scriptscriptstyle K})$ 
can be identified by Eq.~\eqref{eq_DFC_characterize} with the above matrices $(F, G, N)$;
specifically, by substituting $C_{\scriptscriptstyle K} \to SC_{\scriptscriptstyle K}$, 
$B_{\scriptscriptstyle K} \to B_{\scriptscriptstyle K}S$, and 
$A_{\scriptscriptstyle K} \to A_{\scriptscriptstyle K}-B_{\scriptscriptstyle K}C_{\scriptscriptstyle K}$
in Eq.~\eqref{eq_DFC_characterize}, we have 
\begin{align*}
      &SC_{\scriptscriptstyle K}N=F+C,\\
      &B_{\scriptscriptstyle K}S=-N(G+B), \\
      &(A_{\scriptscriptstyle K}
                        -B_{\scriptscriptstyle K}C_{\scriptscriptstyle K})N 
                                 = N(A+BF_{0}+GC),
\end{align*}
which yield 
\begin{align}
\label{controller matrices example 1}
     &A_{\scriptscriptstyle K}=N(A+BF_{0}+GC+G_{0}F_{0})N^{+}, \nonumber \\
     &B_{\scriptscriptstyle K}=-NG_{0}\Sigma, \nonumber \\
     &C_{\scriptscriptstyle K}=\Sigma F_{0}N^{+}, 
\end{align}
where $N^{+}$ is the right inverse to $N$, i.e., $NN^{+}=I_{2}$.

{\bf (iii)}
Note again that the controller \eqref{eq_dynamical_controller} has to satisfy 
the physical realizability condition \eqref{phys real condition}, which is now 
$A_{\scriptscriptstyle K}\Sigma+\Sigma A_{\scriptscriptstyle K}^{\top}
+2B_{\scriptscriptstyle K}\Sigma B_{\scriptscriptstyle K}^{\top}=O$ and 
$B_{\scriptscriptstyle K}=\Sigma C_{\scriptscriptstyle K}^{\top}\Sigma$. 
These constraints are represented in terms of the parameters as follows: \vspace{-1mm}
\begin{align}
\label{controller parameter constraint}
&f_{\scriptscriptstyle 12}=-g_{\scriptscriptstyle 12},~~
f_{\scriptscriptstyle 11}=g_{\scriptscriptstyle 22},~~
n_{\scriptscriptstyle 11}n_{\scriptscriptstyle 22}-n_{\scriptscriptstyle 12}n_{\scriptscriptstyle 21}=-1,~ 
\nonumber \\
&f_{\scriptscriptstyle 12}n_{\scriptscriptstyle 1}=f_{\scriptscriptstyle 11}n_{\scriptscriptstyle 2}-f_{\scriptscriptstyle 14},~~
f_{\scriptscriptstyle 24}+\sqrt{\kappa}=\frac{g}{\sqrt{\kappa}}n_{\scriptscriptstyle 2}, 
\nonumber \\
& \hspace{-0.15cm}
\left( \frac{3}{2}\kappa+\sqrt{\kappa}f_{\scriptscriptstyle 24} \right) n_{\scriptscriptstyle 1}+\omega_{\scriptscriptstyle m}n_{\scriptscriptstyle 2}
=-\sqrt{\kappa}f_{\scriptscriptstyle 11}, 
\nonumber \\
&\omega_{\scriptscriptstyle m}n_{\scriptscriptstyle 1} - \left ( \frac{3}{2}\kappa + \sqrt{\kappa}f_{\scriptscriptstyle 24} \right )n_{\scriptscriptstyle 2}
=\sqrt{\kappa}f_{\scriptscriptstyle 12}, 
\end{align}
where 
$n_{\scriptscriptstyle 1}=n_{\scriptscriptstyle 11}n_{\scriptscriptstyle 24}-
n_{\scriptscriptstyle 14}n_{\scriptscriptstyle 21}$ and 
$n_{\scriptscriptstyle 2}=n_{\scriptscriptstyle 12}n_{\scriptscriptstyle 24}-
n_{\scriptscriptstyle 14}n_{\scriptscriptstyle 22}$. 
This is one of our main results; 
the linear controller \eqref{eq_dynamical_controller} achieving BAE 
for the opto-mechanical oscillator can be fully parametrized by 
Eq.~\eqref{controller matrices example 1} satisfying the condition 
\eqref{controller parameter constraint}. 
We emphasize that this full parametrization of the controller can be 
obtained thanks to the general problem formulation based on the geometric 
control theory.

{\bf (iv)}
In practice, of course, we need to determine a concrete set of parameters to construct the controller. 
Especially here let us consider a {\it passive system}; 
this is a static quantum system such as an empty optical cavity. 
The main reason for choosing a passive system rather than a non-passive 
(or {\it active}) one such as an optical parametric oscillator is that, due to the 
external pumping energy, the latter could become fragile and also its physical 
implementation must be more involved compared to a passive system \cite{Walls2008}. 
Now the condition for the system 
$(A_{\scriptscriptstyle K}, \,B_{\scriptscriptstyle K}, \,C_{\scriptscriptstyle K})$ 
to be passive is given by 
$\Sigma A_{\scriptscriptstyle K}\Sigma=-A_{\scriptscriptstyle K}$ and 
$\Sigma B_{\scriptscriptstyle K}\Sigma=-B_{\scriptscriptstyle K}$; 
the general result of this fact is given in Theorem~3 in Appendix~B. 
From these conditions, the system parameters are imposed to satisfy, in addition 
to Eq.~\eqref{controller parameter constraint}, the following equalities: 
\begin{align}
\label{passive para cond}
    f_{\scriptscriptstyle 12}=\frac{g}{\sqrt{\kappa}}, ~~
    f_{\scriptscriptstyle 11}=0, ~~ 
    n_{\scriptscriptstyle 11}=-n_{\scriptscriptstyle 22}, ~~ 
    n_{\scriptscriptstyle 12}=n_{\scriptscriptstyle 21}.
\end{align}
There is still some freedom in determining $n_{ij}$, which however corresponds 
to simply the phase shift at the input-output ports of the controller, as indicated 
from Eq.~\eqref{controller matrices example 1}. 
Thus, the passive controller achieving BAE in this example is unique up to the 
phase shift. 
Here particularly we chose $n_{\scriptscriptstyle 11}=1$ and 
$n_{\scriptscriptstyle 12}=0$. 
Then the controller matrices \eqref{controller matrices example 1} satisfying 
Eqs.~\eqref{controller parameter constraint} and \eqref{passive para cond} 
are determined as 
\begin{align*}
        A_{\scriptscriptstyle K}
                    =\left[\begin{array}{cc}
                         -\frac{~g^2}{\kappa} &-\omega_{\scriptscriptstyle m} \\
                         \omega_{\scriptscriptstyle m} & -\frac{~g^2}{\kappa}  \\
                      \end{array}\right],~~
        C_{\scriptscriptstyle K}=-B_{\scriptscriptstyle K}^{\top}
                    =\left[\begin{array}{cc}
                         \frac{g}{\sqrt{\kappa}} & 0 \\
                         0 & \frac{g}{\sqrt{\kappa}} \\
                      \end{array}\right]. 
\end{align*}
As illustrated in Fig.~\ref{fig_3I3OBAE}, the controller specified by these 
matrices can be realized as a single-mode, 2-inputs and 2-outputs optical cavity 
with decay rate $g^2/\kappa$ and detuning $-\omega_{\scriptscriptstyle m}$. 
In other words, if we take the cavity with the following Hamiltonian and 
the coupling operator ($\hat{a}_{3}=(\hat{q}_{3}+i \hat{p}_{3})/\sqrt{2}$ is the 
cavity mode) 
\begin{align}
               \hat{H}_{\scriptscriptstyle K}
                             &=\Delta \hat{a}^{\ast}_{3}\hat{a}_{3}
                             =\frac{\Delta}{2}(\hat{q}_{3}^2+\hat{p}_{3}^2),  \nonumber \\
               \hat{L}_{\scriptscriptstyle K}
                             &=\sqrt{\kappa_{\scriptscriptstyle K}}\hat{a}_{3}
                             =\sqrt{\frac{\kappa_{\scriptscriptstyle K}}{2}}
                                        (\hat{q}_{3}+i \hat{p}_{3}), 
\label{eq_controller_H}
\end{align}
then to satisfy the BAE condition the controller parameters 
$(\Delta, \kappa_{\scriptscriptstyle K})$ must satisfy 
\begin{equation}
\label{perfect BAE condition example}
                 \Delta=-\omega_{\scriptscriptstyle m}, ~~~
                 \kappa_{\scriptscriptstyle K}=g^2/\kappa. 
\end{equation}
Summarizing, the above-designed sensing system composed of the opto-mechanical 
oscillator (plant) and the optical cavity (controller), which are combined via coherent 
feedback, satisfies the BAE condition. 
Hence, it can work as a high-precision detector of the force $\hat{f}$ below the SQL, 
particularly when the $\hat P_1$-squeezed probe input field is used; 
this fact will be demonstrated in Section~\ref{sec:5}.


\section{Direct interaction scheme}\label{sec:4}

\begin{figure}[!t]
\begin{center}
\includegraphics[width=8cm,clip]{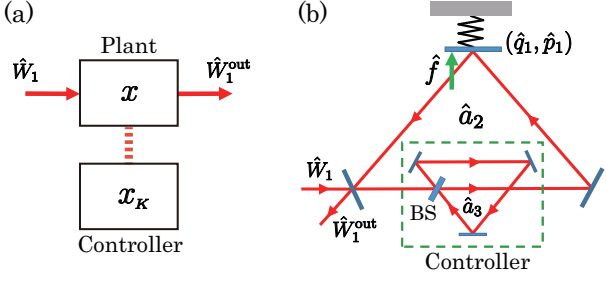}
\caption{(a) General configuration of direct interaction scheme. 
(b) Physical implementation of the passive direct interaction controller for the 
opto-mechanical oscillator.}
\label{fig_Mankei_scheme}
\end{center}
\end{figure}

In this section, we study another control scheme for achieving BAE. 
As illustrated in Fig. \ref{fig_Mankei_scheme}~(a), the controller in this case 
is directly connected to the plant, not through a coherent feedback; 
hence this scheme is called the {\it direct interaction}. 
The controller is characterized by the following two Hamiltonians: 
\begin{align} 
\label{eq_direct_intHamiltonian}
          \hat{H}_{\scriptscriptstyle K}
                 =\frac{1}{2}\hat{x}_{\scriptscriptstyle K}^{\top}
                      R_{\scriptscriptstyle K}\hat{x}_{\scriptscriptstyle K}, ~
          \hat{H}_{\rm int}
                 =\frac{1}{2}(\hat{x}^{\top}R_1\hat{x}_{\scriptscriptstyle K}
                            +\hat{x}_{\scriptscriptstyle K}^{\top}R_2\hat{x}),
\end{align}
where $\hat{x}_{\scriptscriptstyle K}=[\hat{q}^{\scriptscriptstyle \prime}_{1}, 
\hat{p}^{\scriptscriptstyle \prime}_{1}, \ldots , 
\hat{q}^{\scriptscriptstyle \prime}_{n_{\scriptscriptstyle k}}, 
\hat{p}^{\scriptscriptstyle \prime}_{n_{\scriptscriptstyle k}}]^{\top}$ is the 
vector of controller variables with $n_{\scriptscriptstyle k}$ the number of 
modes of the controller. 
$\hat{H}_{\scriptscriptstyle K}$ is the controller's self Hamiltonian with 
$R_{\scriptscriptstyle K}\in 
\mathbb{R}^{2n_{\scriptscriptstyle k} \times 2n_{\scriptscriptstyle k}}$. 
Also $\hat{H}_{\rm int}$ with 
$R_1 \in \mathbb{R}^{2n \times 2n_{\scriptscriptstyle k}}$, 
$R_2 \in \mathbb{R}^{2n_{\scriptscriptstyle k} \times 2n}$ represents the 
coupling between the plant and the controller. 
Note that, for the Hamiltonians $\hat{H}_{\scriptscriptstyle K}$ and 
$\hat{H}_{\rm int}$ to be Hermitian, the matrices must satisfy 
$R_{\scriptscriptstyle K}=R_{\scriptscriptstyle K}^\top$ and $R_1^\top=R_2$; 
these are the physical realizability conditions in the scenario of direct 
interaction. 
In particular, here we consider a plant system interacting with a single probe 
field $\hat W_1$, with coupling matrices $B_1=B$ and $C_1=C$. 
Then, the whole dynamics of the augmented system with variable 
$\hat{x}_{{\scriptscriptstyle E}}=[\hat{x}^{\top}, \hat{x}_{\scriptscriptstyle K}^{\top}]^{\top}$ 
is given by
\begin{align}
\label{direct whole system}
             \frac{d\hat{x}_{{\scriptscriptstyle E}}}{dt}
                       &=A_{{\scriptscriptstyle E}}\hat{x}_{{\scriptscriptstyle E}}
                                            +B_{{\scriptscriptstyle E}}\hat{W}_{1}+b_{{\scriptscriptstyle E}}\hat{f},  \nonumber \\
             \hat{W}_{1}^{\rm out}&=C_{{\scriptscriptstyle E}}\hat{x}_{{\scriptscriptstyle E}}+\hat{W}_{1} ,
\end{align}
where
\begin{align}
\label{eq_direct_matrixAe}
       A_{{\scriptscriptstyle E}}
          &=\left[\begin{array}{cc}
                 ~A ~~& \Sigma_{n}R_1~~ \\
                 \Sigma_{n_{\scriptscriptstyle k}}R_2 
                              & \Sigma_{n_{\scriptscriptstyle k}}R_{\scriptscriptstyle K}\\
            \end{array}\right], ~~
      B_{{\scriptscriptstyle E}}=\left[\begin{array}{c}
                       B\\
                       O \\
                    \end{array}\right],  \nonumber  \\
      C_{{\scriptscriptstyle E}}&=\left[\begin{array}{cc}
                       C & O  \\
                \end{array}\right], ~~
       b_{{\scriptscriptstyle E}}=\left[\begin{array}{cc}
                        b^{\top} & O\\ 
                    \end{array}\right] ^{\top}. 
\end{align}
Note that $B_{{\scriptscriptstyle E}}$, $C_{{\scriptscriptstyle E}}$, and $b_{{\scriptscriptstyle E}}$ 
are the same matrices as those in Eq.~\eqref{extended_matrices}. 
Also, comparing the matrices \eqref{eq_extended_matrixAe} and 
\eqref{eq_direct_matrixAe}, we have that $D_{\scriptscriptstyle K}=O$, which 
thus leads to $F=F_{0}$ and $G=G_{0}$ in Theorem~2. 
Now, again for the opto-mechanical system illustrated in Fig.~\ref{fig_opticalsensor}, 
let us aim to design the direct interaction controller, so that the whole system 
\eqref{direct whole system} achieves BAE; 
that is, the problem is to determine the matrices 
$(R_{\scriptscriptstyle K}, R_1, R_2)$ so that the back-action noise $\hat Q_1$ 
does not appear in the measurement output $\hat P_1^{\rm out}$. 
For this purpose, we go through the same procedure as that taken in Section~\ref{sec:3-3}.

{\bf (i)}
Because of the structure of the matrices $B_{{\scriptscriptstyle E}}$ and $C_{{\scriptscriptstyle E}}$, 
the system is already of the form \eqref{eq_extended_system}, 
where the geometric control theory is directly applicable.

{\bf (ii)}
Because we now focus on the same plant system as that in Section \ref{sec:3-3}, 
the same conclusion is obtained; 
that is, the BAE problem is solvable as long as there is no constraint on the 
controller matrices $(R_{\scriptscriptstyle K}, R_1, R_2)$.

The controller matrices can be determined in a similar way to Section \ref{sec:3-3} as follows. 
First, because the $(C,A,B)$-pair $(\mathcal{V}_{1}, \mathcal{V}_{2})$ is the 
same as before, it follows that ${\rm dim} \, \mathcal{X}_{{\scriptscriptstyle K}}=2$, i.e., 
$n_{\scriptscriptstyle k}=1$. 
Then, from Theorem~2 with the fact that $F=F_{0}$ and $G=G_{0}$, 
we find that the direct interaction controller can be parameterized as follows: 
\begin{align}
\label{direct controller matrices example}
       &R_{\scriptscriptstyle K}=-\Sigma N(A+BF+GC)N^{+}, \nonumber  \\
       &R_1=-\Sigma_2 BFN^{+}, \nonumber  \\
       &R_2=\Sigma NGC,
\end{align}
The matrices $F$, $G$, and $N$ satisfy
$\Kernel F \supseteq \mathcal{V}_{1},  \Image G \subseteq \mathcal{V}_{2}$, and 
$\Kernel N=\mathcal{V}_{1}$, 
which lead to
%
\begin{align}
\label{eq_parameterization2}
\hspace{-0.8cm}
              F&=\left[\begin{array}{cccc}
                        f_{\scriptscriptstyle 11}&f_{\scriptscriptstyle 12}
                               &0&f_{\scriptscriptstyle 14}\\
                        \frac{g}{\sqrt{\kappa}}&0&0&f_{\scriptscriptstyle 24}\\
                     \end{array}\right], ~~
              G=\left[\begin{array}{cc}
                        0&g_{\scriptscriptstyle 12}\\
                        -\frac{g}{\sqrt{\kappa}}&g_{\scriptscriptstyle 22}\\
                        g_{\scriptscriptstyle 31}&g_{\scriptscriptstyle 32}\\
                        0&0\\
                     \end{array}\right], \nonumber  \\
              N&=\left[\begin{array}{cccc}
                        n_{\scriptscriptstyle 11}&n_{\scriptscriptstyle 12}
                              &0&n_{\scriptscriptstyle 14}\\
                        n_{\scriptscriptstyle 21}&n_{\scriptscriptstyle 22}
                              &0&n_{\scriptscriptstyle 24}\\
                     \end{array}\right],
\end{align}
where $f_{ij}, g_{ij}$, and $ n_{ij}$ are free parameters. 

{\bf (iii)}
The controller matrices have to satisfy the physical 
realizability conditions $R_{\scriptscriptstyle K}=R_{\scriptscriptstyle K}^\top$ 
and $R_1^\top=R_2$; 
these constraints impose the parameters to satisfy 
\begin{align}
\label{direct controller parameter constraint}
&f_{\scriptscriptstyle 12}=-g_{\scriptscriptstyle 12},~~
f_{\scriptscriptstyle 11}=g_{\scriptscriptstyle 22}, ~~
n_{\scriptscriptstyle 11}n_{\scriptscriptstyle 22}-n_{\scriptscriptstyle 12}n_{\scriptscriptstyle 21}=-1, \nonumber \\
&f_{\scriptscriptstyle 12}n_{\scriptscriptstyle 1}=f_{\scriptscriptstyle 11}n_{\scriptscriptstyle 2}-f_{\scriptscriptstyle 14}, ~~ 
f_{\scriptscriptstyle 24}=\frac{g}{\sqrt{\kappa}}n_{\scriptscriptstyle 2}, 
\nonumber \\
& \hspace{-0.15cm}
\left( \frac{\kappa}{2}+\sqrt{\kappa}f_{\scriptscriptstyle 24} \right) 
      n_{\scriptscriptstyle 1}+\omega_{\scriptscriptstyle m}n_{\scriptscriptstyle 2}
=-\sqrt{\kappa}f_{\scriptscriptstyle 11}, ~~\nonumber \\
&\omega_{\scriptscriptstyle m}n_{\scriptscriptstyle 1} - \left ( \frac{\kappa}{2} + \sqrt{\kappa}f_{\scriptscriptstyle 24} \right )n_{\scriptscriptstyle 2}
=\sqrt{\kappa}f_{\scriptscriptstyle 12}, 
\end{align}
where 
$n_{\scriptscriptstyle 1}=n_{\scriptscriptstyle 11}n_{\scriptscriptstyle 24}-
n_{\scriptscriptstyle 14}n_{\scriptscriptstyle 21}$ and 
$n_{\scriptscriptstyle 2}=n_{\scriptscriptstyle 12}n_{\scriptscriptstyle 24}-
n_{\scriptscriptstyle 14}n_{\scriptscriptstyle 22}$. 
\\
Equations \eqref{direct controller matrices example}, 
\eqref{eq_parameterization2}, and \eqref{direct controller parameter constraint} 
provide the full parametrization of the direct interaction controller.

{\bf (iv)}
To specify a set of parameters, as in the case of Section \ref{sec:3-3}, let us aim 
to design a passive controller. 
From Theorem~4 in Appendix~B, $R_{\scriptscriptstyle K}$ and 
$R_2=R_1^\top$ satisfy the condition 
$\Sigma R_{\scriptscriptstyle K}\Sigma=-R_{\scriptscriptstyle K}$ and 
$\Sigma R_2 \Sigma_{2}=-R_2$, which lead to the same equalities given in 
Eq.~\eqref{passive para cond}. 
Then, setting the parameters to be $n_{\scriptscriptstyle 11}=1$ and $n_{\scriptscriptstyle 12}=0$, 
we can determine the matrices $R_{\scriptscriptstyle K}$ and $R_2$ as follows: 
\begin{align*}
                R_{\scriptscriptstyle K}
                     =\left[\begin{array}{cc}
                             -\omega_{\scriptscriptstyle m} &0\\
                             0&-\omega_{\scriptscriptstyle m} \\
                       \end{array}\right] , ~~
                R_2=R_1^\top
                   =\left[\begin{array}{cccc}
                             0&0&g&0 \\
                             0&0&0&g \\
                     \end{array}\right] . 
\end{align*}
The controller specified by these matrices can be physically implemented 
as illustrated in Fig.~\ref{fig_Mankei_scheme}~(b); that is, it is a single-mode 
detuned cavity with Hamiltonian $\hat{H}_{\scriptscriptstyle K}
= -\omega_{\scriptscriptstyle m}\hat{a}^{\ast}_{3}\hat{a}_{3}$, which couples 
to the plant through a beam-splitter (BS) represented by
$\hat{H}_{\rm int}=g(\hat{a}_{3}\hat{a}^{*}_{2}+\hat{a}^{*}_{3}\hat{a}_{2})$. 

{\bf Remark:} 
We can employ an active controller, as proposed in \cite{Tsang2010}. 
In this case the interaction Hamiltonian is given by 
$\hat{H}_{\rm int}=g_{\scriptscriptstyle \rm B}(\hat{a}_{3}\hat{a}^{*}_{2}
+\hat{a}^{*}_{3}\hat{a}_{2})+g_{\scriptscriptstyle \rm D}(\hat{a}_{3}\hat{a}_{2}
+\hat{a}^{*}_{3}\hat{a}^{*}_{2})$, 
while the system's self-Hamiltonian is the same as above; 
$\hat{H}_{\scriptscriptstyle K}
= -\omega_{\scriptscriptstyle m}\hat{a}^{\ast}_{3}\hat{a}_{3}$. 
That is, the controller couples to the plant through a non-degenerate optical 
parametric amplification process in addition to the BS interaction. 
To satisfy the BAE condition, the parameters must satisfy 
$g_{\scriptscriptstyle \rm B}+g_{\scriptscriptstyle \rm D}=g$. 
Note that this direct interaction controller can be specified, in the 
full-parameterization \eqref{direct controller matrices example}, 
\eqref{eq_parameterization2}, and \eqref{direct controller parameter constraint}, 
by 
%
%
\[
     f_{\scriptscriptstyle 11}=f_{\scriptscriptstyle 12}=f_{\scriptscriptstyle 14}=0,~~
     n_{\scriptscriptstyle 11}=-n_{\scriptscriptstyle 22}=1,~~ 
     n_{\scriptscriptstyle 12}=n_{\scriptscriptstyle 21}=0.
\]
%


\section{Approximate Back-Action Evasion}\label{sec:5}

We have demonstrated in Sections \ref{sec:3-3} and \ref{sec:4} that the 
BAE condition can be achieved by engineering an appropriate auxiliary 
system and connecting it to the plant. 
However, in a practical situation, it cannot be expected to realize such perfect 
BAE due to several experimental imperfections. 
Hence, in a realistic setup, we should modify our strategy for engineering 
a sensor so that it would accomplish {\it approximate BAE}. 
Then, looking back into Section \ref{sec:2-3} where the BAE condition, 
$\Xi_{Q}(s)=0~\forall s$, was obtained, we are naturally led to consider 
the following optimization problem to design an auxiliary system achieving 
the approximate BAE: 
\begin{align}
\label{eq_approximateBAE}
             \min \Big\|\frac{\Xi_{Q}(s)}{\Xi_{f}(s)}\Big\|, 
\end{align}
where $\|\bullet\|$ denotes a valid norm of a complex function. 
In particular, in the field of robust control theory, the following $H_2$ norm 
and the $H_\infty$ norm are often used \cite{Zhou1996}: 
\[
        \|\Xi\|_2 
         = \sqrt{\frac{1}{2\pi}\int_{-\infty}^{\infty}|\Xi(i\omega )|^{2}d\omega},~~~
        \|\Xi\|_\infty
              = \max_{\omega}|\Xi(i\omega )|. 
\]
That is, the $H_2$ or $H_\infty$ control theory provides a general procedure 
for synthesizing a feedback controller that minimizes the above norm. 
In this paper, we take the $H_2$ norm, mainly owing to the broadband 
noise-reduction nature of the $H_2$ controller. 
Then, rather than pursuing an optimal quantum $H_2$ controller based 
on the quantum $H_2$ control theory \cite{NurdinLQG2009, Hamerly2012}, 
here we take the following geometric-control-theoretical approach to solve the 
problem \eqref{eq_approximateBAE}. 
That is, first we apply the method developed in Section \ref{sec:3} or 
\ref{sec:4} to the idealized system and obtain the controller achieving BAE; 
then, in the practical setup containing some unwanted noise, we make a local 
modification of the controller parameters obtained in the first step, to minimize 
the cost $\|\Xi_{Q}(s)/\Xi_{f}(s)\|_2$.

As a demonstration, here we consider the coherent feedback control for the 
opto-mechanical system studied in Section \ref{sec:2-3}, which is now 
subjected to the thermal noise $\hat{f}_{\rm th}$. 
Following the above-described policy, we employ the coherent feedback 
controller constructed for the idealized system that ignores $\hat{f}_{\rm th}$, 
leading to the controller given by Eqs.~\eqref{eq_controller_H} and 
\eqref{perfect BAE condition example}, illustrated in Fig.~\ref{fig_3I3OBAE}. 
The closed-loop system with variable $\hat{x}_{{\scriptscriptstyle E}}=
[\hat{x}^{\top},\hat{x}_{\scriptscriptstyle K}^{\top}]^{\top}$, 
which now takes into account the realistic imperfections, 
then obeys the following dynamics:
\begin{align}
\label{eq_stateout2}
          \frac{d\hat{x}_{{\scriptscriptstyle E}}}{dt}
                     &=\widetilde{A}_{{\scriptscriptstyle E}}\hat{x}_{{\scriptscriptstyle E}}
                            +B_{{\scriptscriptstyle E}}\hat{W}_{1}                        
                            +b_{{\scriptscriptstyle E}}(\hat{f}_{\rm th}+\hat{f}),  \nonumber \\
          \hat{W}_{3}^{\rm out}
                     &=C_{{\scriptscriptstyle E}}\hat{x}_{{\scriptscriptstyle E}}+\hat{W}_{1}, 
\end{align}
where
\[
            \widetilde{A}_{{\scriptscriptstyle E}}
                 =\left[\begin{array}{cccc|cc}
                        0&\omega_{\scriptscriptstyle m} &0&0&0&0\\
                        -\omega_{\scriptscriptstyle m} &-\gamma &g&0&0&0\\
                        0&0&-\kappa/2 &0&0& \sqrt{\kappa \kappa_{\scriptscriptstyle K}} \\
                        g&0&0&-\kappa/2 &-\sqrt{\kappa \kappa_{\scriptscriptstyle K}}&0\\
                        \hline 0&0&0&\sqrt{\kappa \kappa_{\scriptscriptstyle K}}&0&\Delta \\
                        0&0&-\sqrt{\kappa \kappa_{\scriptscriptstyle K}}&0&-\Delta &0\\
                   \end{array}\right] .
\]
$B_{{\scriptscriptstyle E}}$, $C_{{\scriptscriptstyle E}}$, and $b_{{\scriptscriptstyle E}}$ are the same matrices 
given in Eq.~\eqref{extended_matrices}. 
$\hat{f}_{\rm th}$ is the thermal noise satisfying 
$\langle \hat{f}_{\rm th}(t)\hat{f}_{\rm th}(t') \rangle \simeq \bar{n}\delta (t-t')$,
where $\bar{n}$ is the mean phonon number at thermal equilibrium 
\cite{Giovannetti2001, Wimmer2014}. 
Note that the damping effect appears in the $(2, 2)$ component of 
$\widetilde{A}_{{\scriptscriptstyle E}}$ due to the stochastic nature of $\hat{f}_{\rm th}$.
Also, again, $\kappa_{\scriptscriptstyle K}$ and $\Delta$ are the 
decay rate and the detuning of the controller cavity, respectively. 
In the idealized setting where $\hat{f}_{\rm th}$ is negligible, the perfect BAE is achieved 
by choosing the parameters satisfying Eq.~\eqref{perfect BAE condition example}. 
The measurement output of this closed-loop system is, in the Laplace domain, 
represented by 
\[
         \hat{P}_{3}^{\rm out}(s)
               =\widetilde{\Xi}_{f} (\hat{f}_{\rm th}(s)+\hat{f}(s))
                    +\widetilde{\Xi}_{Q}\hat{Q}_{1}(s)+\widetilde{\Xi}_{P} \hat{P}_{1}(s).
\]
The normalized noise power spectral density of 
$y_{3}(s)=\hat{P}_{3}^{\rm out}(s)/\widetilde{\Xi}_{f}(s)$ is calculated as 
\begin{align}
          \widetilde{S} (\omega ) 
              &=\langle | y_{3}(i\omega )-\hat{f}(i\omega )|^2 \rangle  \nonumber \\
              &=\langle|\hat{f}_{\rm th}|^2 \rangle 
                   + \left| \frac{\widetilde{\Xi}_{Q}}{\widetilde{\Xi}_{f}} \right| ^2 
                                  \langle |\hat{Q}_{1}|^2 \rangle 
                   + \left| \frac{\widetilde{\Xi}_{P}}{\widetilde{\Xi}_{f}} \right| ^2 
                                  \langle |\hat{P}_{1}|^2 \rangle.
\label{eq_noisepower_thermal}
\end{align}
The coefficient of the back-action noise is given by 
\begin{align}
       \frac{\widetilde{\Xi}_{Q}(s)}
                {\widetilde{\Xi}_{f}(s)}=
          -\frac{ \sqrt{\kappa}\{ \kappa \kappa_{\scriptscriptstyle K}\Delta 
                         (s^2+\gamma s + \omega_{\scriptscriptstyle m}^2 )
                             + g^2\omega_{\scriptscriptstyle m} (s^2+\Delta^2 )\} }
                    {g\omega_{\scriptscriptstyle m}\sqrt{\gamma}
                         \{ (s+\kappa/2)(s^2+\Delta^2 ) 
                              +\kappa \kappa_{\scriptscriptstyle K} s\}   }
\label{eq_normalizedtrans_thermal}.
\end{align}
Our goal is to find the optimal parameters $(\kappa_{\scriptscriptstyle K}, \Delta)$ 
that minimize the $H_2$ norm of the transfer function, 
$\widetilde{\Xi}_{Q}/\widetilde{\Xi}_{f}$.

\begin{figure}[!t]
\begin{center}
\includegraphics[width=7cm,clip]{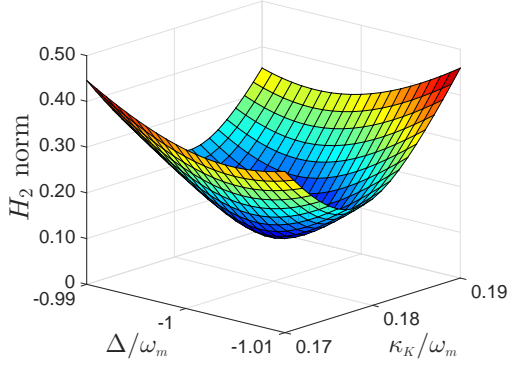}
\caption{
$H_{2}$ norm $\|\widetilde{\Xi}_{Q}/\widetilde{\Xi}_{f}\|_{2}$ 
versus the coupling constant $\kappa_{\scriptscriptstyle K}$ and 
the detuning $\Delta$.}
\label{fig_h2norm}
\end{center}
\end{figure}

\begin{figure}[!t]
\begin{center}
\includegraphics[width=8.5cm,clip]{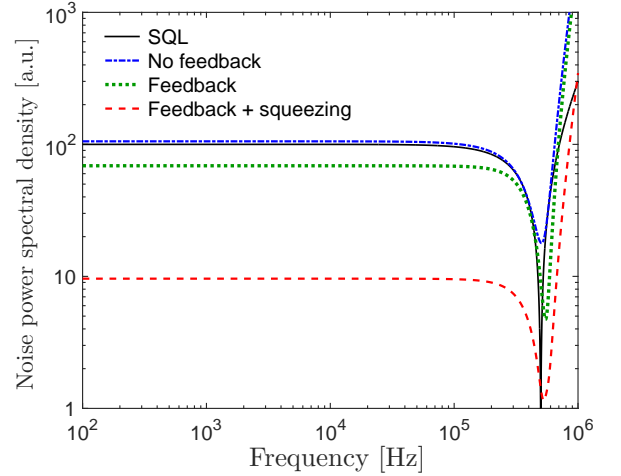}
\caption{Normalized power spectral densities of the noise. 
The black solid line represents the SQL \eqref{eq_SQL_practical}, and 
the dot-dashed blue line does the case without feedback. 
The dotted green and dashed red lines show the cases for the feedback controlled system, 
with coherent and squeezed probe field, respectively. }
\label{fig_noisepower}
\end{center}
\end{figure}

The system parameters are taken as follows \cite{Wimmer2014}: 
$\omega_{\scriptscriptstyle m}/2\pi=0.5$ {\rm MHz}, 
$\kappa/2\pi=1.0$ {\rm MHz}, 
$\gamma/2\pi=5.0$ {\rm kHz}, 
$g/2\pi=0.3$ {\rm MHz}, 
$\bar{n} \simeq 8.33 \times 10^{2}$, 
and the effective mass is $1.0 \times 10^{-12}$ {\rm kg}. 
We then have Fig.~\ref{fig_h2norm}, showing 
$\|\widetilde{\Xi}_{Q}/\widetilde{\Xi}_{f}\|_2$ as a function of 
$\kappa_{\scriptscriptstyle K}$ and $\Delta$. 
This figure shows that there exists a unique pair of 
$(\kappa_{\scriptscriptstyle K}^{\rm opt}, \Delta^{\rm opt})$ that 
minimizes the norm, and they are given by 
$\kappa_{\scriptscriptstyle K}^{\rm opt} /2\pi=0.093$ {\rm MHz} and
$\Delta^{\rm opt}/2\pi=-0.5$ {\rm MHz},
which are actually close to the ideal values \eqref{perfect BAE condition example}. 
Fig.~\ref{fig_noisepower} shows the value of Eq.~\eqref{eq_noisepower_thermal} 
with these optimal parameters 
$(\kappa_{\scriptscriptstyle K}^{\rm opt}, \Delta^{\rm opt})$, where 
the noise floor $\langle|\hat{f}_{\rm th}|^2 \rangle$ is subtracted. 
The solid black line represents the SQL, which is now given by 
\begin{equation}
         \widetilde{S}_{\scriptscriptstyle \rm SQL} (\omega ) 
               =\frac{|(\omega^2 - \omega_{\scriptscriptstyle m}^2)-i\gamma\omega|}
                         {\gamma \omega_{\scriptscriptstyle m}}.
\label{eq_SQL_practical}
\end{equation}
Then the dot-dashed blue and dotted green lines indicate that, in the low frequency range, the 
coherent feedback controller can suppress the noise below the SQL, while, by 
definition, the noise power of the autonomous (i.e., uncontrolled) plant system 
is above the SQL. 
Moreover, this effect can be enhanced by injecting a $\hat{P}_{1}$-squeezed 
probe field (meaning $\langle |\hat{Q}_{1}|^2 \rangle=e^{r}/2$ and 
$\langle |\hat{P}_{1}|^2 \rangle=e^{-r}/2$) into the system. 
In fact the dashed red line in the figure illustrates the case $r=2$ (about 9 dB squeezing), 
showing the significant reduction of the noise power.

\section{Conclusion}
\label{sec:6}

The main contribution of this paper is in that it first provides the general 
theory for constructing a back-action evading sensor for linear quantum 
systems, based on the well-developed classical geometric control theory. 
The power of the theory has been demonstrated by showing that, for the 
typical opto-mechanical oscillator, a full parametrization of the auxiliary 
coherent-feedback and direct interaction controller achieving BAE was 
derived, which contains the result of \cite{Tsang2010}. 
Note that, although we have studied a simple example for the purpose of 
demonstration, the real advantage of the theory developed in this paper 
will appear when dealing with more complicated multi-mode systems such 
as an opto-mechanical system containing a membrane 
\cite{Plenio2008, Meystre2008, Nakamura2016, Nielsen2016}. 
Another contribution of this paper is to provide a general procedure for designing 
an approximate BAE sensor under realistic imperfections; 
that is, an optimal approximate BAE system can be obtained by solving 
the minimization problem of the transfer function from the back-action noise 
to the measurement output. 
While in Section~\ref{sec:5} we have provided a simple approach based on 
the geometric control theory for solving this problem, the $H_2$ or $H_\infty$ control theory could be 
employed for systematic design of an approximate BAE controller even 
for the above-mentioned complicated system. 
This is also an important future research direction of this work.


\appendix


\section*{Appendix A: Algorithms for computing $\mathcal{V}^{*}$ and 
$\mathcal{V}_{*}$}
\label{sec:appendixA}

The set of $(A, B)$-invariant subspaces has a unique maximum element 
contained in a given subspace $\mathcal{H} \subseteq \mathcal{X}$. 
This space, denoted by 
$\mathcal{V}^{*} {\scriptscriptstyle (\mathcal{B}, \mathcal{H})}$, can be 
computed by the following algorithm:
\begin{align*}
\hspace{-0.8cm}\mathcal{V}^{*}~& \mathchar`- ~{\bf algorithm:}~ \\
&{\rm (Step~1)}~\mathcal{V}_{0} := \mathcal{H},\\
&{\rm (Step~2)}~
    \mathcal{V}_{i} 
         := \mathcal{H} \cap A^{-1}(\mathcal{V}_{i-1} \oplus \mathcal{B}) ~~ (i=1, 2, \ldots),\\
&{\rm (Step~3)}~
    \mathcal{V}^{*} {\scriptscriptstyle (\mathcal{B}, \mathcal{H})}
              = \mathcal{V}_{i}~~
                        {\rm (if}~\mathcal{V}_{i}=\mathcal{V}_{i-1}~{\rm in~Step~2)}.
\end{align*}

Similarly, the set of $(C, A)$-invariant subspaces has a unique minimum 
element containing a given subspace $\mathcal{E} \subseteq \mathcal{X}$, 
and this space, denoted by 
$\mathcal{V}_{*} {\scriptscriptstyle (\mathcal{C}, \mathcal{E})}$, can be 
computed by the following algorithm:
\begin{align*}
\hspace{-0.8cm}\mathcal{V}_{*}~& \mathchar`- ~{\bf algorithm:}~ \\
&{\rm (Step~1)}~\mathcal{V}_{0} := \mathcal{E},\\
&{\rm (Step~2)}~
    \mathcal{V}_{i} := \mathcal{E} \oplus  A(\mathcal{V}_{i-1} \cap \mathcal{C}) ~~ (i=1, 2, \ldots),\\
&{\rm (Step~3)}~
           \mathcal{V}_{*} {\scriptscriptstyle (\mathcal{C}, \mathcal{E})}
                 = \mathcal{V}_{i}~~ 
                    {\rm (if}~\mathcal{V}_{i}=\mathcal{V}_{i-1}~{\rm in~Step~2)}.
\end{align*}
%


\section*{Appendix B : Passivity condition of linear quantum systems}
\label{sec:appendixB}

This appendix provides the passivity condition of a general linear quantum system. 
First note that the system dynamics \eqref{eq_LQS_dynamics} and 
\eqref{eq_LQS_output}, which can be represented as 
\begin{align}
\label{eq_QSDE_quadrature}
   \frac{d\hat{x}}{dt}=A \hat{x}+B \hat{W}, ~~~
   \hat{W}^{\rm out}=C\hat{x}+D\hat{W}, 
\end{align}
with $\hat{W}=[\hat{W}_{1}, \ldots, \hat{W}_{m}]^{\top}$, has the following equivalent expression: 
\begin{align}
\hspace{-0.8cm}
   \frac{d}{dt}\left[\begin{array}{c}
                           \hat{a}   \\
                           \hat{a}^{\sharp} \\
          \end{array}\right] 
   &=\mathscr{A}\left[\begin{array}{c}
                                \hat{a}  \\
                                \hat{a}^{\sharp} \\
                             \end{array}\right] 
     + \mathscr{B}\left[\begin{array}{c}
                                \hat{A}  \\
                                \hat{A}^{\sharp} \\
                             \end{array}\right], \nonumber \\
         \left[\begin{array}{c}
                           \hat{A}^{\rm out}  \\
                           \hat{A}^{\rm out}\mbox{}^{\sharp} \\
          \end{array}\right] 
     &=\mathscr{C}\left[\begin{array}{c}
                                \hat{a}  \\
                                \hat{a}^{\sharp} \\
                               \end{array}\right] 
        +\mathscr{D}\left[\begin{array}{c}
                                \hat{A}  \\
                                \hat{A}^{\sharp} \\
                               \end{array}\right] ,
\label{eq_QSDE_annihilation}
\end{align}
where $\hat{a}=[\hat{a}_{1}, \ldots, \hat{a}_{n}]^{\top}$ and 
$\hat{A}=[\hat{A}_{1}, \ldots, \hat{A}_{m}]^{\top}$ are vectors of annihilation operators. 
By definition, $\hat{a}^\sharp=[\hat{a}_{1}^*, \ldots, \hat{a}_{n}^*]^{\top}$. 
The coefficient matrices are of the form
\begin{align}         
    \mathscr{A}&=\left[\begin{array}{cc}
                                \mathscr{A}_{-} & \mathscr{A}_{+}  \\
                                \mathscr{A}_{+}^\sharp & \mathscr{A}_{-}^\sharp \\
                           \end{array}\right] ,~~~
    \mathscr{B}=\left[\begin{array}{cc}
                                \mathscr{B}_{-} & \mathscr{B}_{+}  \\
                                \mathscr{B}_{+}^\sharp & \mathscr{B}_{-}^\sharp \\
                           \end{array}\right] ,  \nonumber \\                        
    \mathscr{C}&=\left[\begin{array}{cc}
                                \mathscr{C}_{-} & ~\mathscr{C}_{+}  \\
                                \mathscr{C}_{+}^\sharp & ~\mathscr{C}_{-}^\sharp \\
                           \end{array}\right] ,~~~
    \mathscr{D}=\left[\begin{array}{cc}
                                \mathscr{D}_{-} & ~ \mathscr{D}_{+}  \\
                                \mathscr{D}_{+}^\sharp & ~ \mathscr{D}_{-}^\sharp \\
                           \end{array}\right].
\label{eq_ABCDmatrix_annihilation}
\end{align}
As in the case of \eqref{eq_QSDE_quadrature}, these matrices have to satisfy 
the physical realizability condition; 
see \cite{GoughPRA2010, PetersenMTNS2010}. 
The passivity condition of this system is defined as follows:

{\bf Definition 5:}~ 
The system \eqref{eq_QSDE_annihilation} is said to be passive if the matrices 
satisfy $\mathscr{A}_{+}=O$ and $\mathscr{B}_{+}=O$, in addition to the 
physical realizability condition.

Note that a passive system is constituted only with 
annihilation operator variables; a typical optical realization of the passive system 
is an empty optical cavity. 
Moreover, $\mathscr{D}_{+}=O$ is already satisfied and $\mathscr{B}_{+}=O$ 
leads to $\mathscr{C}_{+}=O$. 
This is the reason why it is sufficient to consider the constraints only on 
$\mathscr{A}_{+}$ and $\mathscr{B}_{+}$. 
Then the goal here is to represent the conditions 
$\mathscr{A}_{+}=\mathscr{B}_{+}=O$ in terms of the coefficient matrices of 
Eq.~\eqref{eq_QSDE_quadrature}. 
For this purpose, let us introduce the permutation matrix $P_{n}$ as follows; 
for a column vector $z=[z_{1}, z_{2}, \ldots, z_{2n}]^{\top}$, $P_{n}$ is defined 
through $P_{n}z=[z_{1}, z_{3}, \ldots, z_{2n-1}, z_{2}, z_{4}, \ldots, z_{2n}]^{\top}$. 
Note that $P_{n}$ satisfies $P_{n}P_{n}^{\top}=P_{n}^{\top}P_{n}=I_{2n}$. 
Then, the coefficient matrices of the above two system representations are 
connected by 
\begin{align*} 
   A&=P_{n}^{\top}\tilde{\mathscr{A}}P_{n},~~ 
   B=P_{n}^{\top}\tilde{\mathscr{B}}P_{m}, \\
   C&=P_{m}^{\top}\tilde{\mathscr{C}}P_{n},~~ 
   D=P_{m}^{\top}\tilde{\mathscr{D}}P_{m}, 
\end{align*}
where 
\begin{align*} 
\tilde{\mathscr{A}}=\frac{1}{2}
\left[\begin{array}{c}
\mathscr{A}_{-}+\mathscr{A}_{-}^{\sharp}+\mathscr{A}_{+}+\mathscr{A}_{+}^{\sharp} ~~~~~~~\\
-i(\mathscr{A}_{-}-\mathscr{A}_{-}^{\sharp}+\mathscr{A}_{+}-\mathscr{A}_{+}^{\sharp}) ~~~~~~~
\end{array}\right.\\
\left. \begin{array}{c}
~~~~~~~i (\mathscr{A}_{-}-\mathscr{A}_{-}^{\sharp}-\mathscr{A}_{+}+\mathscr{A}_{+}^{\sharp} ) \\
~~~~~~~\mathscr{A}_{-}+\mathscr{A}_{-}^{\sharp}-(\mathscr{A}_{+}+\mathscr{A}_{+}^{\sharp})
\end{array}\right].
\end{align*}
$\tilde{\mathscr{B}}$, $\tilde{\mathscr{C}}$, and $\tilde{\mathscr{D}}$ have the same forms as above. 
Then, we have the following theorem, providing the 
passivity condition in the quadrature form:

{\bf Theorem 3:}~ 
The system \eqref{eq_QSDE_quadrature} is passive if and only if, in addition 
to the physical realizability condition \eqref{phys real condition}, the following 
equalities hold:
\begin{align*}
       \Sigma_{n}A\Sigma_{n}=-A, ~~ \Sigma_{n}B\Sigma_{m}=-B .
\end{align*}
{\it Proof:}~
Let us first define $\Psi_{n}=\Sigma\otimes I_n=[O, I_n ; -I_n, O]$, which leads to 
$P_{n}^{\top}\Psi_{n}P_{n}=\Sigma_{n}$. 
Then, we can prove 
\begin{align*}
     \Sigma_{n}A\Sigma_{n}=-A  ~\Longleftrightarrow~ 
              \Psi_{n}\tilde{\mathscr{A}}\Psi_{n}=-\tilde{\mathscr{A}}. 
\end{align*}
The condition in the right hand side is equivalent to 
$\mathscr{A}_{+}+\mathscr{A}_{+}^{\sharp}=O$ and 
$\mathscr{A}_{+}-\mathscr{A}_{+}^{\sharp}=O$, 
which thus leads to $\mathscr{A}_{+}=O$. Also, from a similar calculation we obtain 
$\Sigma_{n}B\Sigma_{m}=-B \Leftrightarrow \mathscr{B}_{+}=O. 
\hspace{\fill} \blacksquare$

Let us next consider the passivity condition of the direct interaction controller 
discussed in Section~\ref{sec:4}. The setup is that, for a given linear quantum system, 
we add an auxiliary component with variable $\hat{x}_{\scriptscriptstyle K}$, 
which is characterized by the Hamiltonians \eqref{eq_direct_intHamiltonian}. 
The point is that these Hamiltonians have the following equivalent representations
in terms of the vector of annihilation operators $\hat a$ and 
$\hat{a}_{\scriptscriptstyle K}$:
\begin{align} 
\label{eq_direct_intH_annihilation}
          \hat{H}_{\scriptscriptstyle K}&=\frac{1}{2}
                    \left[\begin{array}{cc}
                          \hat{a}_{\scriptscriptstyle K}^{\dag} & \hat{a}_{\scriptscriptstyle K}^{\top} \\
                    \end{array}\right] \mathscr{R}_{\scriptscriptstyle K}
                    \left[\begin{array}{c}
                           \hat{a}_{\scriptscriptstyle K}  \\
                           \hat{a}_{\scriptscriptstyle K}^{\sharp} \\
                    \end{array}\right], \nonumber \\
          \hat{H}_{\rm int}&=\frac{1}{2}\left(
                    \left[\begin{array}{cc}
                          \hat{a}^{\dag} & \hat{a}^{\top} \\
                    \end{array}\right] \mathscr{R}_1
                    \left[\begin{array}{c}
                           \hat{a}_{\scriptscriptstyle K}  \\
                           \hat{a}_{\scriptscriptstyle K}^{\sharp} \\
                    \end{array}\right]
                 +\left[\begin{array}{cc}
                          \hat{a}_{\scriptscriptstyle K}^{\dag} 
                               & \hat{a}_{\scriptscriptstyle K}^{\top} \\
                    \end{array}\right] \mathscr{R}_2
                    \left[\begin{array}{c}
                           \hat{a}  \\
                           \hat{a}^{\sharp} \\
                    \end{array}\right] \right).
\end{align}
The matrices $\mathscr{R}_{\scriptscriptstyle K}, \mathscr{R}_1$, and 
$\mathscr{R}_2$ are of the same forms as those in 
Eq.~\eqref{eq_ABCDmatrix_annihilation}. 
Note that they have to satisfy the physical realizability conditions 
$\mathscr{R}_{\scriptscriptstyle K}=\mathscr{R}_{\scriptscriptstyle K}^\dag$ 
and $\mathscr{R}_1^\dag=\mathscr{R}_2$. 
Now we can define the passivity property of the direct interaction controller; 
that is, if the Hamiltonians \eqref{eq_direct_intH_annihilation} does not 
contain any quadratic term such as $\hat a_{{\scriptscriptstyle K}, 1}^{\ast 2}$ and 
$\hat a_{1}^{\ast}\hat a_{{\scriptscriptstyle K},1}^{\ast}$, then the direct interaction controller 
is passive. 
The formal definition is given as follows:

{\bf Definition 6:}~ 
The direct interaction controller constructed by Hamiltonians 
\eqref{eq_direct_intH_annihilation} is said to be passive if, in addition to the 
physical realizability conditions 
$\mathscr{R}_{\scriptscriptstyle K}=\mathscr{R}_{\scriptscriptstyle K}^\dag$ 
and $\mathscr{R}_1^\dag=\mathscr{R}_2$, the matrices satisfy 
$\mathscr{R}_{{\scriptscriptstyle K}+}=O$ and $\mathscr{R}_{2+}=O$.

Through almost the same way shown above, we obtain 
the following result:

{\bf Theorem 4:}~ 
The direct interaction controller constructed by Hamiltonians 
\eqref{eq_direct_intHamiltonian} is passive if and only if, in addition to the 
physical realizability conditions 
$R_{\scriptscriptstyle K}=R_{\scriptscriptstyle K}^\top$ and 
$R_1^\top=R_2$, the following equalities hold:
\begin{align*}
       \Sigma_{n_{\scriptscriptstyle k}}R_{\scriptscriptstyle K}\Sigma_{n_{\scriptscriptstyle k}}=-R_{\scriptscriptstyle K},
 ~~ \Sigma_{n_{\scriptscriptstyle k}}R_2\Sigma_{n}=-R_2.
\end{align*}
%


\end{document}